\RequirePackage{fix-cm}
\documentclass[twocolumn]{svjour3}       % twocolumn
\smartqed  % flush right qed marks, e.g. at end of proof
\usepackage{graphicx}
\usepackage[numbers, sort&compress]{natbib}
\usepackage{hyperref}
\usepackage{amssymb}
\usepackage{verbatim}
\usepackage{multirow}
\usepackage{threeparttable}
\usepackage{supertabular}
\usepackage{rotating}
\usepackage{rotfloat}
\usepackage{appendix}
\usepackage{makecell}
\usepackage{graphicx}
\usepackage{textcomp}
\usepackage{array} 
\usepackage{longtable}
\usepackage{booktabs}
\usepackage{float}
\usepackage{setspace}
\usepackage[misc]{ifsym}
\hypersetup{
	colorlinks,
linkcolor=blue,
anchorcolor=blue,
citecolor=blue,
urlcolor=blue
}
\usepackage{tabularx}
\usepackage{verbatim}
\usepackage{url}
\usepackage{amsmath}
\usepackage{supertabular} 
\begin{document}

\title{%Insert your title here%\thanks{Grants or other notes
Review of end-to-end speech synthesis technology based on deep learning
}


\author{Zhaoxi Mu     \textsuperscript{1}    \and
        Xinyu Yang \textsuperscript{1,~\Letter}		\and
        Yizhuo Dong \textsuperscript{1}
}


\institute{
	\begin{itemize}
		\item[] {Zhaoxi Mu} \\
		\email{wsmzxxh@stu.xjtu.edu.cn}\\
		\item[\textsuperscript{\Letter}] {Xinyu Yang} \\
		\email{yxyphd@mail.xjtu.edu.cn}\\
		\item[] {Yizhuo Dong} \\
		\email{dyzhuo@stu.xjtu.edu.cn}\\
		\at
		\item[\textsuperscript{1}] Xi'an Jiaotong University, Xi'an, Shaanxi,\\ People’s Republic of China
	\end{itemize}
}

\date{Received: date / Accepted: date}

\maketitle

\begin{abstract}
%Insert your abstract here. Include keywords, PACS and mathematical
%subject classification numbers as needed.
As an indispensable part of modern human-computer interaction system, speech synthesis technology helps users get the output of intelligent machine more easily and intuitively, thus has attracted more and more attention. Due to the limitations of high complexity and low efficiency of traditional speech synthesis technology, the current research focus is the deep learning-based end-to-end speech synthesis technology, which has more powerful modeling ability and a simpler pipeline. It mainly consists of three modules: text front-end, acoustic model, and vocoder. This paper reviews the research status of these three parts, and classifies and compares various methods according to their emphasis. Moreover, this paper also summarizes the open-source speech corpus of English, Chinese and other languages that can be used for speech synthesis tasks, and introduces some commonly used subjective and objective speech quality evaluation method. Finally, some attractive future research directions are pointed out.
\end{abstract}

\section{Introduction}
\label{sec:1}
%Your text comes here. Separate text sections with
With the rapid development of computer science, artificial intelligence, automation and robot control technology, the demand of human-computer interaction has been fully met and the way has become more and more direct and convenient. Human-computer interaction relies heavily on speech communication. The speech system of the machine is divided into three functional modules: voiceprint recognition, speech recognition and speech synthesis. The most difficult and complex task is speech synthesis. This is because that compared to speech and voiceprint recognition, speech synthesis systems usually require more data for training and more complex models for modeling in order to accurately synthesize high-fidelity speech with various styles by inputting simple text.\par
Speech synthesis is also called text-to-speech (TTS) when the input is text. TTS is a frontier technology in the field of information processing, which involves many disciplines such as acoustics, linguistics, and computer science. The main task is to convert input text into output speech. TTS system is the mouth of the intelligent machine, which has been widely used in various fields of people's daily life, such as voice navigation, information broadcast, intelligent assistant, intelligent customer service, and has achieved great economic benefits. Moreover, it is also being applied to some new fields, such as article reading, language education, video dubbing, and rehabilitation therapy. TTS applications has become an important part of people's lives.
\paragraph{Deep learning-based TTS technology} With the development of computer science and technology, the intelligibility and naturalness of synthesized speech have been greatly improved due to the continuous improvement of TTS techniques from the formant-based methods \cite{97vogten1988text,98klatt1987review,99kang2011formant,100khorinphan2014thai,103klatt1980software,107schroder2001emotional} to the unit selection-based waveform cascade methods \cite{1capes2017siri,82chu2003microsoft,101murray1996emotional,104moulines1990pitch,106atal1971speech,197hunt1996unit,271gonzalvo2016recent}, and to the hidden Markov model (HMM)-based statistical parametric speech synthesis (SPSS) methods \cite{29saito2017statistical,30chen2015deep,32nose2016efficient,83tokuda2013speech,102kawahara2008tandem,105yoshimura1999simultaneous,108zen2007hmm,196zen2009statistical}. Deep learning is a new research direction in the field of artificial intelligence in recent years. This method can effectively capture the latent information and association in data, and has more powerful modeling ability than traditional statistical learning methods \cite{110yang2014deep}. TTS methods based on deep learning have been widely researched \cite{109ze2013statistical,198fernandez2013f0,199lu2013combining,200qian2014training}. For example, in the SPSS model based on deep neural network (DNN), DNN can learn the mapping function from linguistic features (input) to acoustic features (output).\par
DNN-based acoustic models provide an effective distributed representation of the complex dependencies between linguistic features and acoustic features. However, one limitation of the acoustic feature modeling method based on feedforward DNN is that it ignores the continuity of speech. The DNN-based method assumes that each frame is sampled independently, although there is correlation between consecutive frames in the speech data. Recurrent Neural Network (RNN) provides an effective method to model the correlation between adjacent frames of speech, because it can use all the available input features to predict the output features of each frame. Based on this, some researchers use RNN instead of DNN to capture the long-term dependence of speech frames in order to improve the quality of synthesized speech \cite{195zen2015unidirectional,201tuerk1993speech,202karaali1998text,203fan2014tts,204fernandez2014session,205zen2014statistical}.
\paragraph{End-to-end TTS technology}The traditional SPSS network is a complex pipeline containing many modules, composed of text-to-phoneme network, audio segmentation network, phoneme duration prediction network, fundamental frequency prediction network and vocoder \cite{3arik2017deep,4gibiansky2017deep}. Building these modules requires a lot of professional knowledge and complex engineering implementation, which will take a lot of time and effort. Also, the combination of errors in each component may make the model difficult to train. End-to-end TTS methods are driven by the desire to simplify TTS systems and reduce the need for manual intervention and linguistic background knowledge. The end-to-end TTS model only needs to be trained from scratch on the paired data set of $\langle$text, speech$\rangle$, and can directly synthesize speech from the text. The state-of-the-art end-to-end TTS models based on deep learning have been able to synthesize speech close to human voice \cite{2wang2017tacotron,5shen2018natural,17oord2016wavenet}. \par
\begin{figure}
	\includegraphics[width=0.48\textwidth]{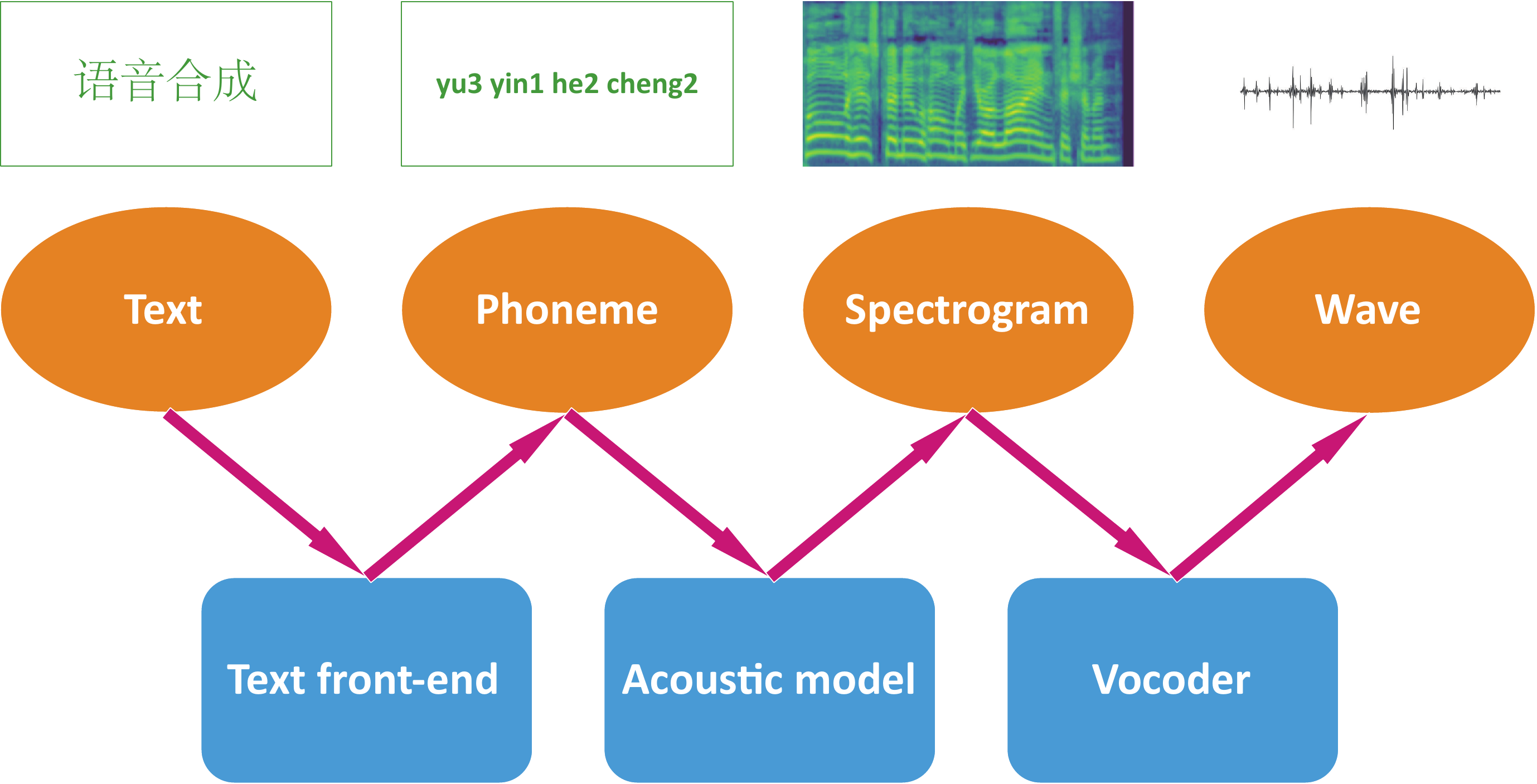}
	\caption{Pipeline architecture for TTS}
	\label{fig:1}       % Give a unique label
\end{figure}
\begin{figure*}
	\includegraphics[width=0.7\textwidth]{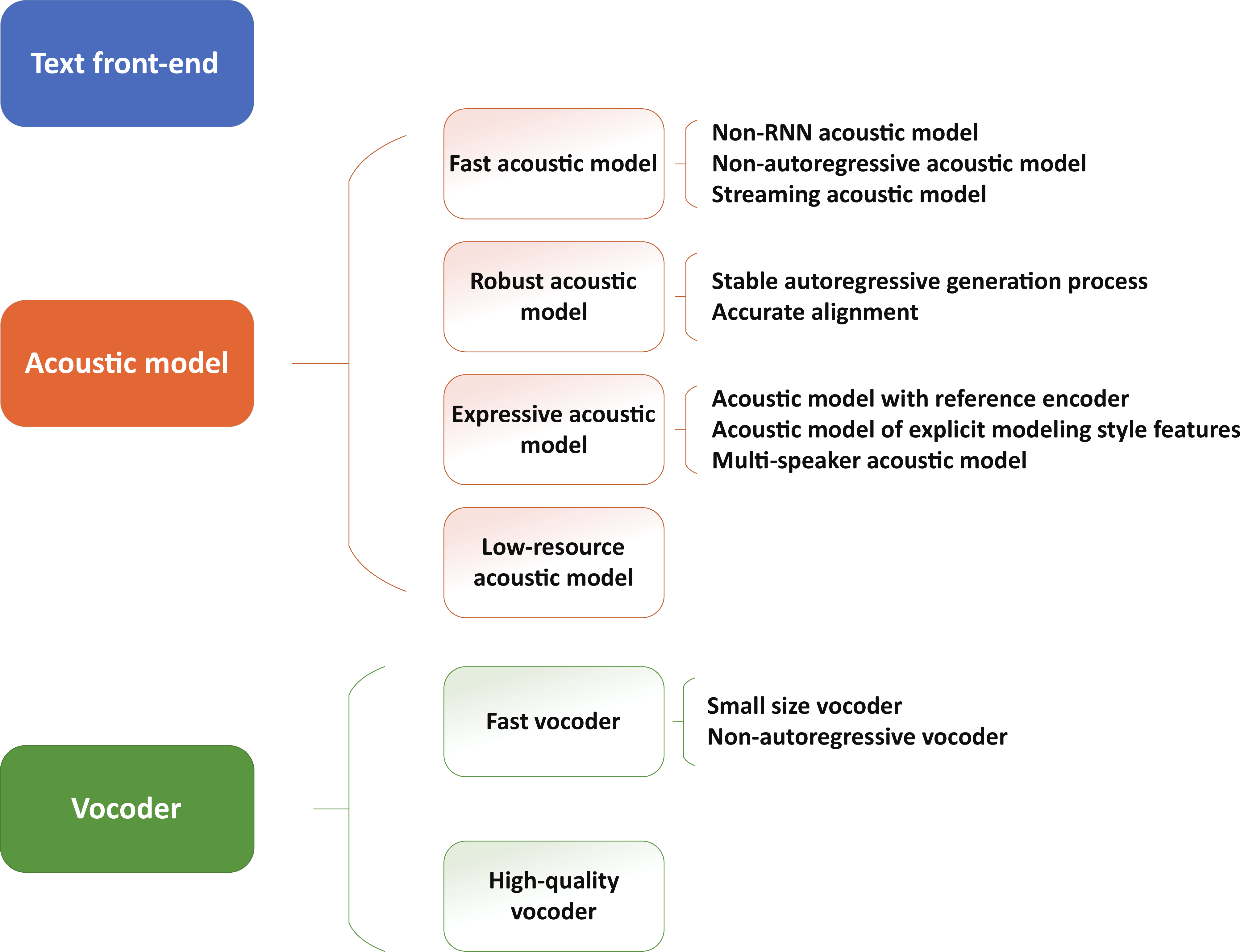}
	\caption{Section organization of the TTS model}
	\label{fig:2}       % Give a unique label
\end{figure*}

It is mainly composed of three parts: text analysis front-end, acoustic model and vocoder, as shown in Fig. \ref{fig:1}. Firstly, the text front-end converts the text into standard input. Then, the acoustic model converts the standard input into intermediate acoustic features, which are used to model the long-term structure of speech. The most common intermediate acoustic features are spectrogram \cite{2wang2017tacotron,5shen2018natural}, vocoder feature \cite{129sotelo2017char2wav} or linguistic feature \cite{17oord2016wavenet}. Finally, the vocoder is used to fill in low-level signal details and convert acoustic features into time-domain waveform samples. To reduce the difficulty of training and improve the quality of synthesized speech, the text front-end, acoustic model and vocoder are usually trained separately \cite{5shen2018natural}, and they can also be fine-tuned jointly \cite{129sotelo2017char2wav}. This article will introduce some of the latest developments in each of the three components according to the structure of Fig. \ref{fig:2}.\par

There have been some reviews on TTS. For example, \citet{147deng2018historical} analyzed the number of documents and citations of TTS papers from 1992 to 2017, aiming to help researchers understand the development trend of TTS. \citet{148aroon2015statistical} reviewed SPSS methods based on HMM. \citet{149adiga2019acoustic} reviewed SPSS methods and partially deep learning based methods. \citet{150ning2019review} and \citet{151sruthi2020review} reviewed TTS methods based on deep learning. \citet{152kalita2017emotional} reviewed emotional TTS methods for Hindi. \citet{153tits2019emotional} reviewed the emotional speech corpus that could be used for TTS.\par
Although there have been some reviews on TTS methods based on deep learning, only some of baseline models have been introduced, such as WaveNet \cite{17oord2016wavenet}, Tacotron \cite{2wang2017tacotron} and SampleRNN \cite{18mehri2016samplernn}. These models have many problems, such as slow training and inference speed, instability, lack of emotion and rhythm in synthesized speech, and a large amount of high-quality speech data required for training. The state-of-the-art TTS methods can completely or partially solve these problems, still so far there has been no comprehensive review of the latest deep learning-based TTS models. Moreover, the quantity and quality of training speech corpus play a decisive role in the training results of TTS model, and how to effectively evaluate the quality of synthesized speech has always been a problem in the field of TTS. Therefore, this paper will make a detailed summary of the latest end-to-end TTS models based on deep learning, speech corpus and evaluation methods of synthesized speech, and finally give some future research directions.\par
The rest of this paper is organized as follows: Sect. \ref{sec:2}, \ref{sec:3} and \ref{sec:4} respectively introduce the latest text front-end, acoustic model and vocoder based on deep learning. Sect. \ref{sec:5} organizes the corpus that could be used for TTS. Sect. \ref{sec:6} introduces commonly used synthesized speech evaluation methods from both subjective and objective aspects. Sect. \ref{sec:7} puts forward some challenges and future research directions for reference. The last section draws a general conclusion of this paper.

\section{Text front-end}
\label{sec:2}
%Text with citations \cite{RefB} and \cite{RefJ}.
It is difficult to synthesize high-fidelity speech only using original phonemes or original text as the input of the TTS model, especially for languages that contain polyphonic characters and have complex prosodic structures, such as Mandarin. Therefore, it is necessary to use the text front-end to introduce additional pronunciation and syntactic information. The text front-end predicts the pronunciation mode from the original text, aiming to provide enough information for the back-end to accurately synthesize speech. The quality of the text front-end has a great impact on the clarity and naturalness of the synthesized speech. Pronunciation patterns are important information for languages with many polyphonic characters and ambiguous pronunciations, such as Mandarin. Syntactic information also contributes a lot to the pronunciation of a sentence, which determines the pause and tone of a sentence. People usually read a phrase that has a full meaning in its entirety, and pause between phrases that need to be separated. For languages with many ambiguities, the effect of syntactic information on sentence segmentation may also cause listeners to have a completely different understanding of a sentence. Therefore, this information needs to be predicted by the text front-end as a conditional input of the acoustic model to synthesize speech with correct pronunciation and prosody.\par
The traditional Mandarin text front-end is a cascade system, which consists of a series of text processing components, such as text normalization (TN), Chinese word segmentation (CWS), part-of-speech (POS) tagging, grapheme-to-phoneme (G2P) and prosodic structure prediction (PSP). The text front-end structure of other languages is similar to that of Mandarin. These components are usually modeled by traditional statistical methods, such as syntactic trees \cite{255zhang2016mandarin} and CRF \cite{258qian2010automatic} based methods for PSP tasks and dictionary matching based methods \cite{256huang2010disambiguation} for pronunciation prediction tasks. However, these traditional text front-ends often fail to predict correctly in some unusual or complex contexts. To boost prediction accuracy, some researchers have adopted state-of-the-art NLP frameworks based on deep learning methods such as BLSTM-CRF \cite{88zheng2018blstm,92huang2015bidirectional}, Word2Vec \cite{90mikolov2013efficient}, Transformer \cite{11vaswani2017attention} and BERT \cite{56devlin2018bert} to improve the text front-end model based on dictionary and traditional statistical learning methods. These models can extract contextual information from the text effectively, and thus help the text front-end to accurately determine the pronunciation of polyphonic characters, the meaning of ambiguous sentences, and the prosodic boundaries between each word, each phrase and each sentence. The following will introduce the latest text front-end model based on deep learning from the aspects of text normalization, prosodic structure prediction, pronunciation prediction, contextual information extraction and so on.
\paragraph{Text normalization} Text normalization is an important preprocessing step for TTS tasks. \citet{75zhang2020hybrid} standardized Mandarin text by combining the traditional rule-based system with a neural text network consisting of multi-head self-attention modules in Transformer to convert Non-Standard Words (NSW) into Spoken-Form Words (SFW). This method has a higher prediction accuracy than the rule-based system.
\paragraph{Prosodic structure prediction} Prosodic structure prediction is also an important function of the text front-end. Taking Mandarin as an example, the prosodic structure of Mandarin is a three-level hierarchical structure composed of three basic units: prosodic words (PW), prosodic phrases (PPH) and intonation phrases (IPH) \cite{96chu2001locating}. Because these three levels of prediction tasks are interrelated, \citet{76pan2019mandarin} modeled prosody information at all levels of the text in the way of multi-task learning, and proposed a Mandarin prosodic boundary prediction model based on BLSTM-CRF, which improved the prediction accuracy and simplified the model. \citet{257lu2019self} also proposed a method of multi-tasking learning to efficiently complete PSP tasks based on the self-attention model.
\paragraph{Pronunciation prediction} Other text front-ends have the pronunciation prediction function on the basis of text normalization and prosody prediction. The G2P tasks of Mandarin can be divided into two categories: G2P of monophonic characters and G2P of polyphonic characters. The pronunciation of monophonic characters can be easily determined by a pronunciation dictionary, while G2P of polyphonic characters is highly context sensitive \cite{261zhang2020unified}. Therefore, disambiguation of polyphonic characters is the main task of Mandarin G2P. To accurately predict the pronunciation of polyphonic characters, \citet{259cai2019polyphone}, \citet{260shan2016bi} and \citet{263park2020g2pm} proposed to use Bi-LSTM network for G2P. On the basis of \citet{76pan2019mandarin}, \citet{77yang2019pre} proposed to preprocess the original text by replacing the Word2Vec model with the encoder of Transformer-based NLP model and BERT pre-training model, and then carry out G2P and PSP in the Mandarin text front-end. The accuracy of prediction can be improved by taking advantage of Transformer and BERT network. However, pre-training models, such as BERT, are too large to be used in realtime applications and edge devices. To reduce the size of the model, \citet{261zhang2020unified} proposed to use the simplified TinyBERT model \cite{262jiao2019tinybert} for the G2P and PSP tasks simultaneously using multi-task learning. It can ensure the accuracy of the prediction results while reducing the size of the model. \citet{78conkie2020scalable} proposed a text front-end that can be used to process multiple languages, including text normalization and G2P functions. They regard these two front-end tasks as two neural machine translation (NMT) tasks and use Transformer for modeling. Byte pair encoding (BPE) technology \cite{89sennrich2015neural} is also used to process uncommon words, and the splicing technique is used for long texts, which improves the accuracy of prediction and the quality of synthesized speech.
\paragraph{Introduction of style information} The text front-end can also directly add additional style information to the TTS system to provide the synthesized speech style features. For example, \citet{45tahon2018can} added a pronunciation adaptive framework based on CRF between text front-end and TTS model to generate different styles of speech. In order to make the synthesized speech closer to human voice, \citet{111szekely2020breathing} took the front and back utterances of an utterance and the breath pronunciation events between them as a data set to learn the breath location information of the context, thus adding human breath information into the training data. The forward and backward breath predictors were also used to predict the location of breath more accurately.
\paragraph{Contextual information extraction} The text front-end model can also extract the contextual information of the text. The extracted additional contextual information can be input into the acoustic model as prior knowledge. For example, \citet{55hayashi2019pre} directly used BERT as a context feature extraction network to encode input text, and added encoded word or sentence-level contextual information to the input of the encoder of the acoustic model to improve the quality of synthesized speech. In order to obtain the phrase structure of the sentence and word relationship information, \citet{54guo2019exploiting} used the factor parser \cite{91klein2003fast} in the Stanford parser to extract the syntactic tree. Then, the embedding vectors of extracted syntactic features and input tokens are then combined as the input of the acoustic model encoder, enabling TTS models to correctly synthesize speech when facing some ambiguous sentences. In order to improve the quality of synthesized speech, GraphSpeech \cite{193liu2020graphspeech} inputs syntactic knowledge as additional contextual information into the self-attention module of Transformer-TTS \cite{10li2019neural}. The syntax tree of the input text is converted into a syntax graph to model the language relation between any two characters in the input text, describe the global relation between the input characters and extract grammatical features of the text.
\paragraph{Unified text front-end} To reduce the cumulative training error of each part and simplify the model, the components of the text front-end with various functions can be combined together. \citet{79pan2020unified} proposed a Mandarin text front-end model that unifies a series of text processing components, which can directly convert the original text into linguistic features. Firstly, the original text is normalized by the method proposed by \citet{75zhang2020hybrid} Then, the Word2Vec model is used to convert sentences into character embedding, and an auxiliary model composed of dilated convolution or Transformer encoder is used to predict CWS and POS respectively. Finally, the results are embedded and combined with the original characters as the input of the main module to jointly predict the labels of phoneme, tone and prosody.

\section{Acoustic model}
\label{sec:3}
Tacotron \cite{2wang2017tacotron} is the first end-to-end acoustic model based on deep learning, and it is also the most widely used acoustic model. It can synthesize acoustic features directly from text, and then synthesize speech waveforms according to Griffin-Lim algorithm \cite{115griffin1984signal}. Tacotron is based on the Seq2Seq architecture of encoder-decoder with attention mechanism. The encoder is composed of the CBHG network and is used to encode the input text. The CBHG network includes convolution bank, highway networks and Bi-GRU \cite{84chung2014empirical}. Decoder consists of RNN with attention mechanism that aligns the output of the encoder with the mel-spectrogram to be generated. Finally, the decoder maps the output sequence of the encoder to the mel-spectrogram in an autoregressive manner \cite{49van2016pixel}. The autoregressive generative method is to decompose the joint probability $p(\boldsymbol{x})$ of the acoustic feature sequence $\boldsymbol{x}=\{x_1,x_2,\ldots,x_T\}$ into:
\begin{equation}
	p(\boldsymbol{x})=\prod_{i=0}^{T-1}p(x_{i+1}\mid x_{1},x_{2},\ldots,x_{i})
\end{equation}
This means that the acoustic features of the $n$-th frame are generated under the condition of the previous $n-1$ frames. In order to increase the speed of synthesizing mel-spectrogram, Tacotron generates multiple frames of mel-spectrogram at each decoding step.\par 
Although Tacotron is better than most SPSS models, it still has the following four disadvantages:
\begin{itemize} 
\item[$\bullet$] The decoder in Tacotron is composed of RNN and synthesizes acoustic features in an autoregressive manner, which introduces a time-series dependence. Therefore, it cannot be calculated in parallel, resulting in slow training and inference speed.
\item[$\bullet$] Tacotron uses content-based attention mechanism, thus the synthesized speech will have many errors, such as mispronunciation, missed words and repetitions.
\item[$\bullet$] Tacotron cannot synthesize speech with a specific emotion and rhythm.
\item[$\bullet$] Tacotron needs to use a lot of high-fidelity speech data during training to get good results.
\end{itemize}
\par
In order to overcome these disadvantages in Tacotron, researchers have proposed many new acoustic models based on Tacotron. The following will introduce various improvement methods for the above four disadvantages.
\subsection{Fast acoustic model}
Although Tacotron can synthesize high-fidelity speech that is close to human voice, it cannot be used in practical applications due to its slow training and inference speed. The training and inference speed of acoustic model can be improved by improving RNN network, improving autoregressive generative method and using streaming method.

\subsubsection{Non-RNN acoustic model}
Multi-layer CNN can replace RNN to capture the long-term dependence of the context, and can speed up training and inference in the way of parallel computing. For example, Tacotron 2 \cite{5shen2018natural} replaces the complex CBHG and GRU structures with simple LSTM \cite{85hochreiter1997long} and CNN structures on the basis of Tacotron. Deep Voice 3 \cite{6ping2017deep} uses residual gated convolution \cite{7dauphin2017language,8gehring2017convolutional} instead of RNN to capture contextual information, where the encoder and decoder are composed of non-causal and causal CNNs. DCTTS \cite{9tachibana2018efficiently} replaces RNN with CNN on the basis of Tacotron, which consists of Text2Mel and Spectrogram Super Resolution Network (SSRN). \par
In addition to CNN, other networks can be used instead of RNN to achieve parallel computing. For example, \citet{10li2019neural} proposed to use Transformer to replace the RNN and attention networks in Tacotron 2, thereby increasing the computational efficiency by using the multi-head self-attention in Transformer to generate the hidden states of encoder and decoder in parallel. \citet{12bi2018deep} proposed that the deep feed-forward sequential memory network (DFSMN) \cite{86zhang2018deep} with a structure similar to dilated-CNN \cite{17oord2016wavenet} can be used to replace RNN in the acoustic model. The quality of speech generation by the DFSMN-based model is similar to that of the RNN-based model, and the model complexity is reduced and the training time is reduced.
\subsubsection{Non-autoregressive acoustic model}
\label{sec:3.1.2}
Although the above models improve the computational efficiency by means of parallel computation, they still need to generate acoustic features frame by frame in an autoregressive manner \cite{49van2016pixel} during inference, resulting in a very slow generation speed. Therefore, if acoustic features can be generated in parallel, the generation speed will be greatly improved. However, it is difficult for the acoustic model based on the attention mechanism to learn the correct alignment between input and output if the mel-spectrogram is directly generated in parallel in a non-autoregressive manner. In order to solve this problem, FastSpeech \cite{13ren2019fastspeech}, SpeedySpeech \cite{165vainer2020speedyspeech}, ParaNet \cite{120peng2020non}, FastPitch \cite{112lancucki2020fastpitch} and other models introduced a teacher network to replace the implicit autoregressive alignment method of the traditional seq2seq model through knowledge distillation. The autoregressive teacher network can guide the non-autoregressive network to learn correct attention alignment.\par
FastSpeech consists of the feed-forward Transformer networks, which can generate acoustic feature frames in parallel under the guidance of the length regulator. The length regulator aligns each language unit with a corresponding number of acoustic frames in a manner provided by the autoregressive teacher network. However, the Transformer module is complex and has a large number of parameters. To reduce model parameters and further improve the speed of training and inference, DeviceTTS \cite{192huang2020devicetts}, SpeedySpeech \cite{165vainer2020speedyspeech}, TalkNet \cite{136beliaev2020talknet}, and Parallel Tacotron \cite{207elias2020parallel} replace the Transformer module in FastSpeech with simple DFSMN \cite{86zhang2018deep}, residual dilated-CNN, CNN and lightweight convolution (LConv) \cite{208wu2019pay}, respectively. \par
The training process for models such as Fastspeech, Speedyspeech, and Paranet is complicated by the use of knowledge distillation. To simplify the training process, other generative models such as normalizing flow and generative adversarial network (GAN) generative models can be used to avoid autoregressive generation and knowledge distillation process. Glow-TTS \cite{141kim2020glow} uses the Glow \cite{22kingma2018glow} normalizing flow instead of Transformer as the decoder to generate mel-spectrogram in parallel (the Glow normalizing flow will be described in detail in Sect. \ref{sec:4.1.2}). Flow-TTS \cite{179miao2020flow} also uses a Glow-based decoder to generate mel-spectrogram non-autoregressively. \citet{114donahue2020end} proposed an end-to-end TTS model EATS based on GAN-TTS \cite{113binkowski2019high}, which directly synthesized speech non-autoregressively using GAN. Table \ref{tab:1} lists the methods to improve training and inference speed of each model.\par
\subsubsection{Streaming acoustic model}
Although the training and inference speed of TTS models has been greatly improved, most of the current models can only output speech after inputting an entire sentence. The longer the sentence, the longer the waiting time, that is, the system will delay the input, which seriously affects the experience of human-computer interaction experience. To solve this problem, some researchers have proposed streaming incremental TTS systems \cite{246yanagita2019neural,247stephenson2020future,248ellinas2020high,249ma2019incremental}, which can output speech in real time while inputting text, because they only need to see a few characters or words to synthesize speech. The streaming system can generate new audio while the user plays the audio, which greatly improves the applicability of the TTS system and the user experience. It can be applied in the fields of simultaneous translation, dialog generation, and assistive technologies \cite{249ma2019incremental}.\par
Traditional acoustic models with complete sentences as input can rely on the full linguistic context (ie, past and future words) to construct their internal representations for acoustic features, thus generating high-quality speech. However, due to the limited contextual information that streaming acoustic models can obtain, it is a challenge to effectively model the overall prosodic structure of speech. \citet{246yanagita2019neural} proposed the streaming neural TTS model for the first time. In order to learn the intra-sentence boundary features, they used the start, middle and end symbols to split the training sentence into multiple subunits, which were used to train the Tacotron. And they allow the model to learn the acoustic time-series within one full sentence by taking the last vector of the mel-spectrogram from the previous units as the initial input for each unit. Finally, the entire sentence is synthesized by incrementally synthesizing blocks consisting of one or more words with symbols. \par
This method needs to preprocess the training data, and only considers the previous information, which will cause the prosodic error of synthesized speech. In order to solve this problem, \citet{249ma2019incremental} borrowed the idea of prefix-to-prefix framework of simultaneous machine translation \cite{251ma2018stacl}. When generating acoustic features and speech waveforms incrementally, not only the previous results but also the information of the following words should be be used as the condition. \citet{247stephenson2020future} also proposed that the following words should be considered when incrementally encoding each word. They use Bi-LSTM to encode the first word to the following few words of the word to be synthesized, and then input the resulting embedding vector into the decoder. Finally, the speech segments will be cropped \cite{250kisler2017multilingual} and spliced. \citet{248ellinas2020high} proposed a streaming inference method, which can input the generated acoustic frames into the vocoder before the inference process of the acoustic model is completed. They accumulate the output frames from each decoding step in a buffer, and when the buffer includes enough frames to accommodate the total receptive field of the convolutional layers in post-net, the acoustic frames are passed to post-net in a larger batch. The post-net is trained to refine the entire acoustic frames sequence. The acoustic frames in the buffer are partially redundant to consider the contextual information of the acoustic frame to be synthesized. \citet{266stephenson2021alternate} used the language model GPT-2 \cite{267radford2019language} to predict the next word in the input text, thereby improving the naturalness of speech synthesized by the incremental TTS model by utilizing the predicted contextual information.
\begin{table*}
	\small
	\caption{Methods to improve the training and inference speed of each acoustic model}
	\label{tab:1}       % Give a unique label
\begin{spacing}{1.6}
	\begin{tabular}{|p{4cm}|p{2cm}|p{3cm}|p{6cm}|}
		\hline\noalign{\smallskip}
		Acoustic model                         & Neural network types   & Generative model types                  & Characteristics                                                                                     \\ \noalign{\smallskip}\hline\noalign{\smallskip}
		Tacotron (\citeauthor{2wang2017tacotron}, \citeyear{2wang2017tacotron})           & CBHG, GRU              & Autoregression                          & Synthesizing speech end-to-end, the structure is complex, the training and inference speed is slow  \\
		Deep Voice 3 (\citeauthor{6ping2017deep}, \citeyear{6ping2017deep})        & CNN                    & Autoregression                          & Based on CNN, training and inference speed is faster than Tacotron                                  \\
		DCTTS (\citeauthor{9tachibana2018efficiently}, \citeyear{9tachibana2018efficiently})         & CNN                    & Autoregression                          & Based on CNN, training and inference speed is faster than Tacotron                                  \\
		Tacotron 2 (\citeauthor{5shen2018natural}, \citeyear{5shen2018natural})         & LSTM, CNN              & Autoregression                          & The structure is simpler than Tacotron                                                              \\
		Transformer-TTS (\citeauthor{10li2019neural}, \citeyear{10li2019neural})      & Transformer            & Autoregression                          & Based on Transformer, training and inference speed is faster than Tacotron                          \\
		FastSpeech (\citeauthor{13ren2019fastspeech}, \citeyear{13ren2019fastspeech})          & Transformer            & Non-autoregression                      & Training through knowledge distillation, training speed is slow, inference speed is fast            \\
		ParaNet (\citeauthor{120peng2020non}, \citeyear{120peng2020non})            & CNN                    & Non-autoregression                      & Training through knowledge distillation, based on CNN, the structure is simpler than FastSpeech     \\
		EATS (\citeauthor{114donahue2020end}, \citeyear{114donahue2020end})            & CNN                    & GAN                 & Based on CNN and GAN, the training and inference speed is fast, the structure is fully end-to-end   \\
		Glow-TTS (\citeauthor{141kim2020glow}, \citeyear{141kim2020glow})            & Transformer, Glow      & Normalizing flow    & Based on normalizing flow, training and inference speed is fast                                     \\
		SpeedySpeech (\citeauthor{165vainer2020speedyspeech}, \citeyear{165vainer2020speedyspeech})     & CNN                    & Non-autoregression                      & Training through knowledge distillation, based on CNN, the structure is simpler than FastSpeech     \\
		TalkNet (\citeauthor{136beliaev2020talknet}, \citeyear{136beliaev2020talknet})         & CNN                    & Non-autoregression                      & Based on CNN, training and inference speed is faster, the structure is simpler than FastSpeech      \\
		Flow-TTS  (\citeauthor{179miao2020flow}, \citeyear{179miao2020flow})           & Glow                   & Normalizing flow    & Based on normalizing flow, training and inference speed is fast                                     \\
		DeviceTTS (\citeauthor{192huang2020devicetts}, \citeyear{192huang2020devicetts})         & DFSMN, RNN             & Combination of autoregression and non-autoregression   & Based on DFSMN, the structure is simpler than FastSpeech                                            \\
		Parallel Tacotron (\citeauthor{207elias2020parallel}, \citeyear{207elias2020parallel}) & LConv                  & Non-autoregression                      & Based on LConv, the structure is simpler than FastSpeech                                            \\
		FastPitch (\citeauthor{112lancucki2020fastpitch}, \citeyear{112lancucki2020fastpitch}) & Transformer			& Non-autoregression					& Training through knowledge distillation, training speed is slow and inference speed is fast						\\	
		\noalign{\smallskip}\hline
	\end{tabular}
\end{spacing}
\end{table*}
\subsection{Robust acoustic model}
The neural TTS models based on autoregressive generative method and attention mechanism have been able to generate speech that is as natural as human voice. However, these models are not as robust as traditional methods. During training, the autoregression-based models need to first decide whether it should stop when predicting each frame. Therefore, incorrect prediction of a single frame can result in serious errors, such as ending the the generation process early. Moreover, there are almost no constraints in the attention mechanism of the acoustic model to prevent problems such as repetition, skipping, long pauses, or nonsense. These errors are rare and therefore usually do not show up in small test sets such as those used in subjective listening tests. However, in customer-oriented products, even if there is only a small probability of such problems, it will greatly reduce the user experience. Therefore, many improved methods for autoregressive generative model and attention mechanism widely used in neural TTS models have been proposed.
\subsubsection{Stable autoregressive generation process}
In order to improve the training convergence speed, the autoregressive TTS models such as Tacotron use natural acoustic feature frames as the input of decoder for teacher forcing training in training stage, while in inference stage, use the previously predicted acoustic feature frames as the input of the decoder to generate speech in free running mode. The distribution of the data predicted by the model is different from the distribution of the real data used in the training process, and the discrepancy between these two distributions can quickly accumulate errors in decoding, resulting in exposure bias and wrong results, such as skipping, repeating words, incomplete synthesis and inappropriate prosody phrase breaks. And this makes the model can only be used to synthesize short sentences, because the sound quality will deteriorate as the length of the synthesized sentence increases.\par
A simple method to reduce exposure bias is scheduled sampling \cite{80bengio2015scheduled}, in which acoustic feature frames of the current time step are predicted by using natural acoustic feature frames or those predicted by the previous time step with a certain probability \cite{79pan2020unified,166morrison2020controllable}. However, due to the inconsistency between the natural speech frames and the predicted speech frames during the scheduled sampling, the temporal correlation of the acoustic feature sequence is destroyed, leading to the decline of the quality of the synthesized speech. \par
To avoid this problem, \citet{133guo2019new} proposed to use the Professor Forcing \cite{265lamb2016professor} method for training, which is a GAN-based adversarial training method. The model is composed of a generator and a discriminator. The generator generates the output sequence in the manner of teacher forcing and free running, respectively. The discriminator based on self-attention GAN (SAGAN) \citet{155zhang2019self} is used to determine which way the output sequence is generated. They reduce the exposure bias by introducing an additional term to minimize the gap between the output sequences generated by the two methods in the training goal of the generator, although this solution is not stable and easy enough. \citet{26liu2019new} proposed the random descent method, which first uses the natural acoustic features as the input of the decoder for the first round of teacher forcing training, and then replaces the natural acoustic features with the acoustic features generated in the first round for the second round of teacher forcing training. The model is trained multiple iterations to minimize the gap between the generated acoustic features and the natural acoustic features, thereby reducing the exposure bias. \citet{81liu2020teacher} also proposed a method based on knowledge distillation to reduce exposure bias, which is to train a teacher model first, and then use it to guide the training of the student model. The teacher model uses ground-truth data for training, and the student model uses the predicted value of the previous time step to guide the prediction of the next time step. Knowledge distillation is performed by minimizing the distance between the hidden states of the decoder at each time step of the two models.\par
When the target sequence is generated by autoregressive method, the previous wrong token will affect the next one. The acoustic feature sequence is usually longer than the target sequence of other sequence learning tasks (such as NMT). Therefore, the results of the TTS task will be more susceptible to error propagation, resulting in that the right part of the generated acoustic feature sequence is usually worse than the left part. \citet{68ren2019almost} used the bidirectional sequence modeling (BSM) technique to alleviate error propagation. They generated acoustic feature sequences from left to right and from right to left respectively to prevent the model from generating sequences with poor quality on one side. \citet{134zheng2019forward} proposed two BSM methods for acoustic models, which take full advantage of the autoregressive model at the initial iteration stage and reduce errors in synthesized speech by adding bidirectional decoding regularization term to the loss function during training. The first method is to construct two acoustic models that generate the mel-spectrogram from front to back and from back to front respectively, and then minimize the difference between the output mel-spectrogram of the two models. The second method is to use two decoders to generate mel-spectrogram forward and backward while sharing an encoder, and then minimize the difference between the state or attention weight values of the two decoders at each time step. Moreover, \citet{165vainer2020speedyspeech} employed three data augmentations on the input mel-spectrogram to improve the robustness of the model to error propagation during autoregressive generation:
\begin{itemize} 
	\item[$\bullet$] A small amount of Gaussian noise is added to each spectrogram pixel.
	\item[$\bullet$] The model outputs are simulated by feeding the input spectrogram through the network without gradient update in parallel mode.
	\item[$\bullet$] The input spectrograms are degraded by randomly replacing several frames with random frames, thereby encouraging the model to use temporally more distant frames.
\end{itemize}
\par
When acoustic features are generated by autoregressive acoustic models, there is a problem of local information preference \cite{213chen2016variational,214liu2019maximizing}, that is, the acoustic feature frames to be generated by the current time step are completely dependent on the acoustic feature frames generated by the previous time step, and are independent of the text conditions. In order to avoid ignoring text information during synthesis and thus generating wrong speech, \citet{214liu2019maximizing} learned from the idea of InfoGAN \cite{215chen2016infogan} and proposed to use an additional auxiliary CTC recognizer to recognize the predicted acoustic features. The predicted acoustic features are used to restore the corresponding input text. This method essentially maximizes the mutual information between the predicted acoustic features and the input text to enhance the dependence between them.
\subsubsection{Accurate alignment}
\begin{figure}
	\includegraphics[width=0.48\textwidth]{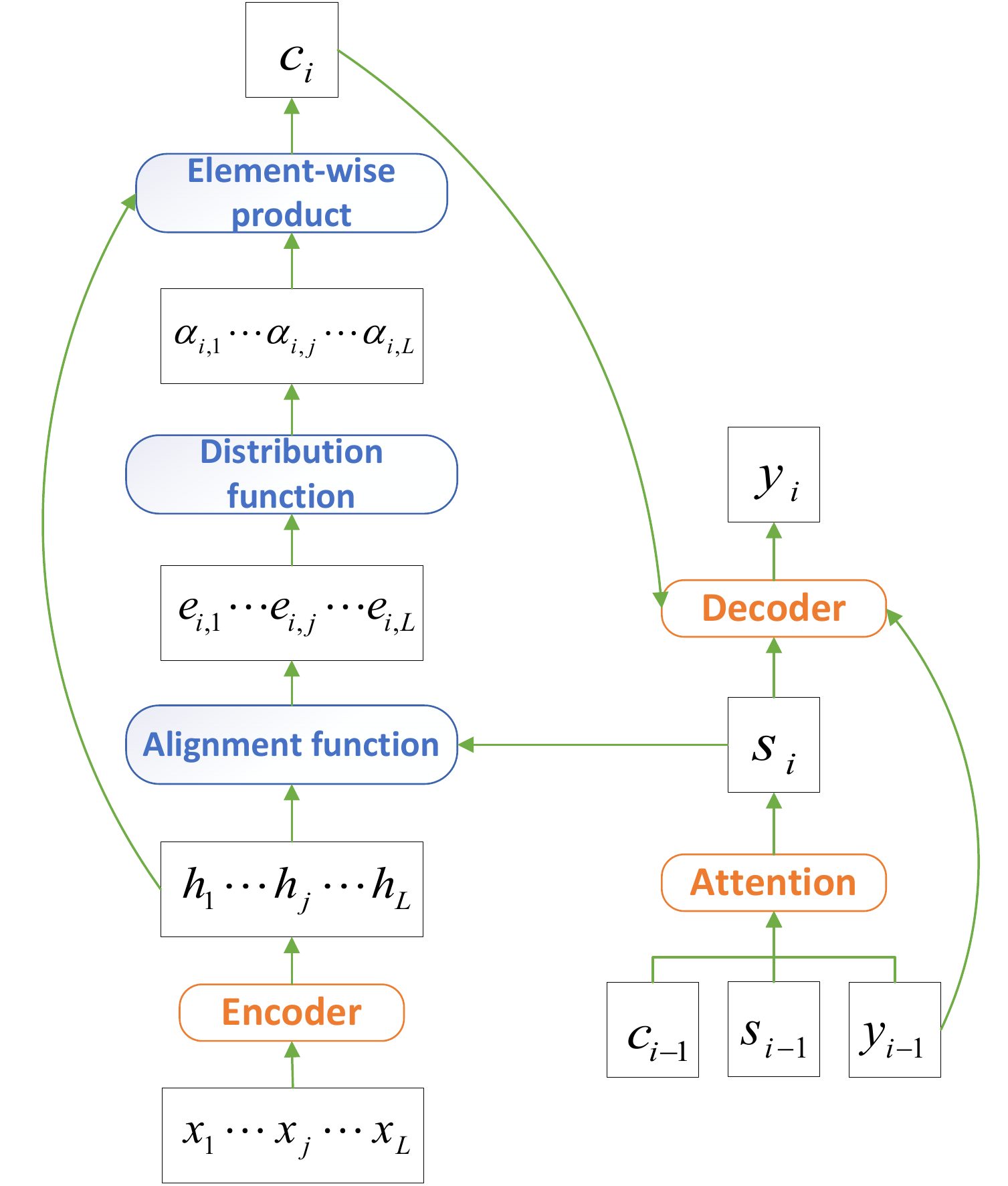}
	\caption{Attention mechanism structure}
	\label{fig:3}       % Give a unique label
\end{figure}

Similar to other Seq2Seq models, many TTS models use the attention mechanism to align input text with output spectrograms. The attention mechanism allows the output of the decoder at each step to focus on a subset of hidden states of the encoder, and the result directly controls the duration and rhythm of the synthesized speech. The main structure of the attention mechanism is shown in Fig. \ref{fig:3}, which can be expressed as \cite{170chaudhari2019attentive}:
\begin{equation}
	(h_1,h_2,\ldots,h_L)=Encoder(x_1,x_2,\ldots,x_L)
	\label{eq.2}
\end{equation}
\begin{equation}
	s_i=Attention(s_{i-1},c_{i-1},y_{i-1})
	\label{eq.3}
\end{equation}
\begin{equation}
	e_{i,j}=f_a (s_i,h_j)
	\label{eq.4}
\end{equation}
\begin{equation}
	\alpha_{i,j}=f_d(e_{i,j})
	\label{eq.5}
\end{equation}
\begin{equation}
	c_i=\sum_j\alpha_{i,j}h_j
	\label{eq.6}
\end{equation}
\begin{equation}
	y_i=Decoder(y_{i-1},c_i,s_i)
	\label{eq.7}
\end{equation}
where $\{x_j\}_{j=1}^L$ is input sequence, $L$ is the length of input sequence, $\{h_j\}_{j=1}^L$ are hidden states of encoder, $c_i$ is context vector, $\alpha_{i,j}$ are attention weights over input, $s_i$ is hidden state of decoder, $e_{i,j}$ are energy values, $y_i$ is output token, $f_a$ is alignment function, $f_d$ is distribution function, and the form of $f_a$ and $f_d$ depends on the specific attention mechanism.
\par
First, the input sequence $(x_1,x_2,\ldots,x_L)$ is encoded by encoder and transformed to $(h_1,h_2,\ldots,h_L)$. Then, the hidden states $\{s_i\}_{i=1}^T$ of decoder is generated by the attention network, and the corresponding weights $\{\alpha_{i,j}\}_{j=1}^L$ of encoder states in the $i$-th time step are calculated by $s_i$. The context vector $c_i$ consists of a linear combination of attention weights $\{\alpha_{i,j}\}_{j=1}^L$ and encoder states $\{h_j\}_{j=1}^L$. Finally, the decoder generates the output token $y_i$ using the current time step context vector $c_i$ and hidden state $s_i$.\par
Since the order and position of input text and output speech in TTS task are corresponding, attention alignment in TTS is a surjective mapping from the output frames to the input tokens and should follow such strict criteria \cite{142he2019robust}:
\begin{itemize} 
	\item[$\bullet$]\emph{Locality} Each output frame should be aligned around a single input token to avoid attention collapse.
	\item[$\bullet$]\emph{Monotonicity} The position of the aligned input token must never rewind backward to prevent repeating.
	\item[$\bullet$]\emph{Completeness} Each input token should be covered once or aligned with some output frame to avoid skipping.
\end{itemize}
\par
The original Tacotron model uses the content-based attention mechanism proposed by \citet{60bahdanau2014neural}. In this case, Eq. (\ref{eq.4}) is:
\begin{equation}
	e_{i,j}=v^Ttanh(Ws_i+Vh_j+b)
	\label{eq.8}
\end{equation}
where $Ws_i$ and $Vh_j$ represent query and key, respectively. 
\par
The content-based attention mechanism does not consider the position information of each item in the sequence at all, and can not effectively utilize the monotonicity and locality of alignment, thus alignment errors are common. In order to enable the attention mechanism to consider the positon information of input and output, and thus enhance the generalization ability of synthesizing long sentences, Char2wav \cite{129sotelo2017char2wav}, Voiceloop \cite{128taigman2017voiceloop} and Melnet \cite{252vasquez2019melnet} adopted the Gaussian mixture model (GMM) attention mechanism proposed by \citet{140graves2013generating} to replace the content-based attention mechanism in Tacotron. This method is a purely location-based attention mechanism, which uses an unnormalized mixture of $K$ Gaussians to produce the attention weights, $\alpha_{i,j}$, for each encoder state:
\begin{equation}
	\alpha_{i,j}=\sum_{k=1}^K\frac{w_{i,k}}{Z_{i,k}}exp\Big(-\frac{(j-\mu_{i,k})^2}{2(\sigma_{i,k})^2}\Big)
	\label{eq.9}
\end{equation}
\begin{equation}
	\mu_{i,k}=\mu_{i-1,k}+\Delta_{i,k}
	\label{eq.10}
\end{equation}
where $w_{i,k}$, $Z_{i,k}$, $\Delta_{i,k}$ and $\sigma_{i,k}$ are computed from the attention RNN state. The mean of each Gaussian component $\mu_{i,k}$ is computed using the recurrence relation in Eq. (\ref{eq.10}), which makes the mechanism location-relative and potentially monotonic if $\Delta_{i,k}$ is constrained to be positive. Although this location-based attention mechanism can enhance the generalization ability of acoustic models for long sentences, it sacrifices some of the naturalness of synthesized speech.
\par
In order to combine content and location information in alignment, Tacotron 2 uses the hybrid location-sensitive attention mechanism \cite{61chorowski2015attention}. In this case, Eq. (\ref{eq.4}) is:
\begin{equation}
	e_{i,j}=v^Ttanh(Ws_i+Vh_j+Uf_{i,j}+b)
	\label{eq.11}
\end{equation}
where $Uf_{i,j}$ represents the location-sensitive term, and uses convolutional features computed from the previous attention weights $\{\alpha_{i-1,j}\}_{j=1}^L$. This method combines the content and location features to make alignment more accurate by additionally introducing previous attention weight information.
\par
Based on the monotonicity of alignment between input and output sequences in TTS, various monotonic attention mechanisms have been proposed to reduce errors in attention alignment. In order to introduce monotonicity into the hybrid location-sensitive attention, \citet{59battenberg2020location} proposed Dynamic Convolution Attention (DCA), which removed content-based terms $Ws_i$ and $Vh_j$, leaving only location-sensitive term $Uf_{i,j}$ as static filters, while adding a set of learned dynamic filters $Tg_{i,j}$ and a single fixed prior filter $p_{i,j}$. In this case, Eq. (\ref{eq.4}) is redefined as:
\begin{equation}
	e_{i,j}=v^Ttanh(Uf_{i,j}+Tg_{i,j}+b)+p_{i,j}
	\label{eq.12}
\end{equation}
Similar to static filters $Uf_{i,j}$, dynamic filters $Tg_{i,j}$ are computed from the attention RNN state and serve to dynamically adjust the alignment relative to the alignment at the previous step. Prior filter $p_{i,j}$ is used to bias the alignment toward short forward steps. This monotonic DCA has stronger generalization ability and is more stable.
\par
\citet{95raffel2017online} proposed a monotonic alignment method that can be applied to TTS: monotonic attention (MA). At each step $i$, MA inspects the memory entries from the memory index $t_{i-1}$ it focused on at the previous step and evaluates the "selection probability" $p_{i,j}$:
\begin{equation}
	p_{i,j}=\sigma({e_{i,j}})
	\label{eq.13}
\end{equation}
where $\sigma$ is logistic sigmoid function and energy values $e_{i,j}$ are produced as in Eq. (\ref{eq.4}). Starting from $j=t_{i-1}$, at each time MA would sample $z_{i,j}\sim Bernoulli(p_{i,j})$ to decide to keepj unmoved $(z_{i,j}=1)$ or move to the next position $(z_{i,j}=0)$. j would keep moving forward until reaching the end of inputs, or until receiving a positive sampling result $z_{i,j}=1$, and when j stops, memory $h_j$ would be directly picked as $c_i$. With such restriction, it is guaranteed that solely one input unit would be focused on at each step, and its position would never rewind backward. Moreover, the mechanism only requires linear time complexity and supports online inputs, which could be efficient in practice.\par
In contrast to the traditional “soft” attention using continuous weights, MA, which simply selects one input unit as the context vector $c_i$, is a “hard” attention. It can ensure the locality of attention alignment, but it could not be trained by standard back-propagation (BP) algorithm. Multiple approaches have been proposed for this issue, including reinforcement learning \cite{143xu2015show,144zaremba2015reinforcement,145ling2017coarse}, approximation by beam search \cite{146shankar2018surprisingly}, and approximation by soft attention for training \cite{95raffel2017online}.\par
\begin{figure*}
	\includegraphics[width=0.85\textwidth]{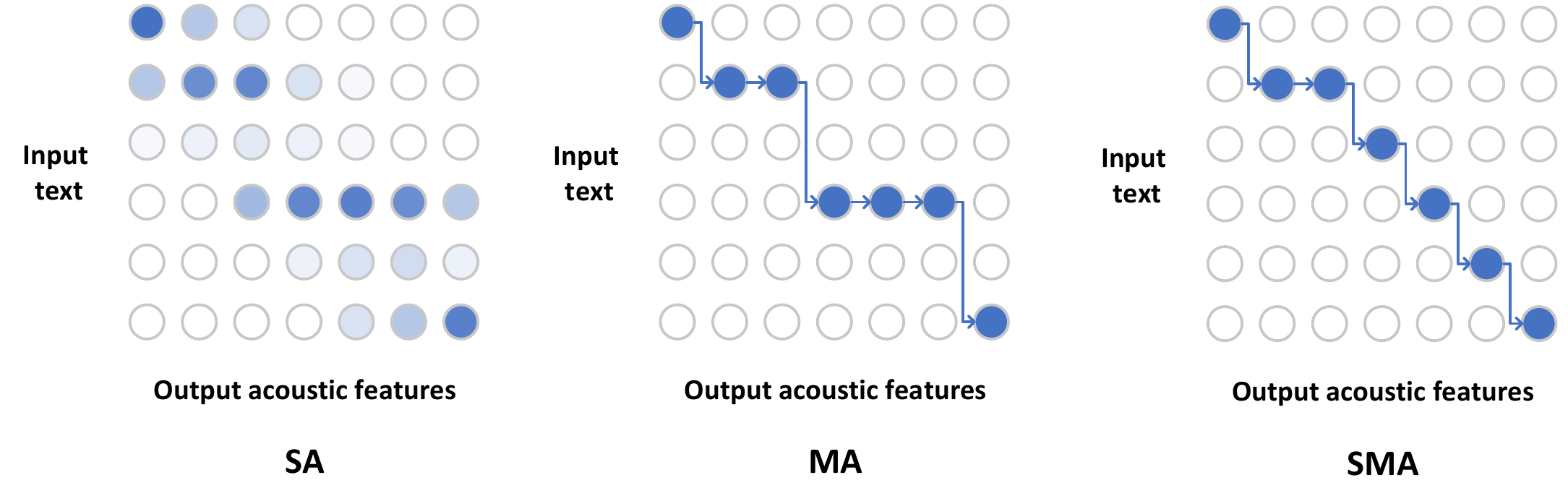}
	\caption{The alignment of SA, MA and SMA}
	\label{fig:4}       % Give a unique label
\end{figure*}

To further guarantee the completeness of alignment, \citet{142he2019robust} proposed stepwise monotonic attention (SMA), which adds additional restrictions on MA: in each decoding step, the attention alignment position moves forward at most one step, and it is not allowed to skip any input unit. The alignment of soft attention (SA), MA and SMA is shown in Fig. \ref{fig:4} \cite{142he2019robust}. The color depth of each node in the figure represents the size of the attention weight between each output acoustic feature frame and the input phoneme. The darker the color, the greater the value of attention weight. The figure shows that each acoustic feature frame is calculated by multiple input phonemes in SA. Each acoustic feature frame is determined by an input phoneme in MA. In SMA, not only each acoustic feature frame is determined by an input phoneme, but all input phonemes must be corresponding at least once, which ensures the locality, monotonicity and completeness of attention alignment.\par
\citet{174zhang2018forward} and \citet{177yasuda2019initial} also proposed similar monotonic attention mechanisms. \citet{174zhang2018forward} suggested that only the alignment paths satisfying the monotonic condition are taken into consideration at each decoder time step. The attention probabilities of each time step can be computed recursively using a forward algorithm, and a transition agent is proposed to help the attention mechanism make decisions whether to move forward or stay at each decoder time step. This attention mechanism has the advantages of fast convergence speed and high stability. \citet{177yasuda2019initial} also proposed a hard monotonic attention mechanism. The framework and likelihood function are similar to those of a hidden Markov model (HMM). The constrained alignment is conceptually borrowed from segment-to-segment neural transduction (SSNT) \cite{175yu2016online,176yu2016neural}. They factorized the generation probability for acoustic features into an alignment transition probability and emission probability, thereby constraining the alignment process to moving from left to right, and only one step at a time. Although this hard monotonic alignment method can avoid some alignment errors that are commonly observed in soft-attention-based methods, including muffling, skipping, and repeating, this attention mechanism has poor stability and long training time.\par
In order to make more direct use of the correspondence between text and speech in TTS, \citet{9tachibana2018efficiently} and \citet{16wang2019deep} added a guided attention loss to content-based dot product attention \cite{11vaswani2017attention}. More specifically, they added an additional monotonic attention loss to the original audio reconstruction loss, forcing the non-zero values of the attention weight matrix were concentrated on the diagonal as much as possible. Furthermore, the forced increment attention was proposed to force the text and speech to be aligned monotonously by making the corresponding text position of acoustic feature frame at each time step move forward by at most one. To produce monotonic alignment, Deep Voice 3 and ParaNet added positional encoding in Transformer to the content-based dot product attention. Besides, they added an attention window \cite{26liu2019new,27merboldt2019analysis} he attention during inference, calculated the attention weights only for the input characters in the window, and took the position of the character with the largest attention weight as the starting position of the next window. Moreover, ParaNet adopted a multi-layer attention mechanism to iteratively refine attention alignment in a layer-by-layer manner.\par
However, the use of positional encoding can cause errors when synthesizing long sentences \cite{253li2020robutrans}. To synthesize long sentences stably, Glow-TTS removes the positional encoding and adds relative position representations \cite{154shaw2018self} into self-attention modules instead. RobuTrans \cite{253li2020robutrans} counts on the 1-D CNN used in Encoder Pre-net to model relative position information in a fixed window. Moreover, in order to make the self-attention in Transformer more suitable for TTS models, RobuTrans also uses Pseudo Non-causal Attention (PNCA) to replace the traditional causal self-attention. The decoding process is more robust by providing the decoder with the holistic view of the input sequence and the frame-level context information.\par
As described in Sect. \ref{sec:3.1.2}, a large number of non-autoregressive acoustic models have been proposed recently. TTS is a one-to-many mapping. For the same text input, there are many possible speech expressions with different prosody. To eliminate ambiguity in multi-mode output, the acoustic models with autoregressive decoders can predict the acoustic feature frames of the next time step by combining the contextual information provided by the acoustic feature frames generated by the previous time step. However, acoustic models with non-autoregressive decoders need to obtain contextual information in other ways to select an appropriate generation mode. Non-autoregressive acoustic models need to determine the output length in advance, rather than predict whether to stop at each frame. In this case, in order to align the inputs and outputs, a duration predictor similar to the one used in the traditional SPSS method \cite{109ze2013statistical,196zen2009statistical} can be used instead of the attention network. Aligning with a duration predictor can avoid the errors of skipping, repeating, and irregular stops caused by the attention mechanism. This method first appeared in NMT \cite{14gu2018meta}, and then was introduced into TTS through non-autoregressive acoustic models such as FastSpeech \cite{13ren2019fastspeech}. Acoustic models with duration predictors can align input phonemes and output acoustic features by introducing additional alignment modules or using external aligners. Next, these two alignment methods are introduced separately.\par
The most direct way to obtain the alignment information is provided by an external aligner. For example, FastSpeech extracts phoneme duration from a pre-trained autoregressive model by knowledge distillation \cite{268kim2016sequence}. However, FastSpeech lacks generalization ability for long utterances, especially those whose length exceeds the maximum length of the utterance in the training set. This may be because the self-attention is a global modeling method. To use the local modeling method to make network more stable, DeviceTTS \cite{192huang2020devicetts} replaces the Transformer with DFSMN, which makes use of a latency control window size to learn the context. To simplify the training process, JDI-T \cite{137lim2020jdi} jointly trains the autoregressive Transformer teacher network and the feed-forward Transformer student network. To avoid the complicated knowledge distillation process, some models use a separate external alignment model to predict the target phoneme duration, thus establishing alignment between input phonemes and output acoustic features. For example, TalkNet \cite{136beliaev2020talknet} uses the CTC-based automatic speech recognition (ASR) model Quartznet \cite{178kriman2020quartznet}, FastSpeech 2 \cite{15ren2020fastspeech} uses the forced-alignment tool MFA toolkit \cite{87mcauliffe2017montreal}, DurIAN \cite{66yu2019durian} uses a external alignment model \cite{169zen2016fast,203fan2014tts}, RobuTrans \cite{253li2020robutrans} uses speech recognition tools, Parallel Tacotron \cite{207elias2020parallel} and Non-Attentive Tacotron \cite{138shen2020non} use a speaker-dependent HMM-based aligner with a lexicon \cite{73wightman1997aligner}. To address the difficulty of training an aligner due to data sparsity, \citet{138shen2020non} used fine-grained VAE (FVAE) to achieve semi-supervised and unsupervised duration prediction, that is, simply train the model using the predicted durations instead of the target durations for upsampling.\par
It is also possible to directly learn alignment by training an alignment module within the model. For example, AlignTTS \cite{135zeng2020aligntts} uses the dynamic programming to consider all possible alignments in training, that is, uses the alignment loss inspired by the Baum-Welch algorithm \cite{158taylor2009text,159baum1970maximization} to train the mix density network for alignment. Glow-TTS uses the Monotonic Alignment Search (MAS) algorithm to predict the duration of each input tokens by searching for the most probable monotonic alignment between text and the latent representation of speech. The internal alligator of EATS \cite{114donahue2020end} implicitly enhances the monotonicity of alignment by predicting token lengths and obtaining positions using a cumulative sum operation. Moreover, the dynamic time warping (DTW) loss and the aligner length loss are introduced to learn alignment and ensure that the model can accurately predict phoneme lengths. Flow-TTS \cite{179miao2020flow} trains a length predictor inside the model to predict the output length in advance, and takes the positional encoding of the predicted spectrogram length as query vector to align the input and output using the positional attention module based on the multi-head dot-product attention mechanism \cite{11vaswani2017attention}.\par
Since one-to-many regression problems like TTS can benefit from autoregressive decoding, it is also possible to combine the autoregressive method with duration predictor to further improve the stability of TTS models, such as the alignment methods used in DurIAN, Non-Attentive Tacotron \cite{138shen2020non}, DeviceTTS and RobuTrans \cite{253li2020robutrans}. The alignment method of each model is shown in Table \ref{tab:2}.
\begin{table*}
	\small
	\caption{Alignment method of each acoustic model}
	\label{tab:2}       % Give a unique label
	\begin{spacing}{1.2}
		\begin{tabular}{|p{2.5cm}|p{1.8cm}|p{2.5cm}|p{4cm}|p{4cm}|}
					\hline\noalign{\smallskip}
		Acoustic model                             & Neural network types & Generative model types                & Alignment methods                                                                                             & Characteristics                                                                       \\
		\noalign{\smallskip}\hline\noalign{\smallskip}
		Tacotron (\citeauthor{2wang2017tacotron}, \citeyear{2wang2017tacotron})               & CBHG, GRU            & Autoregression                          & Content-based attention                                                                                       & Unstable, alignment errors often occur                                              \\
		Char2Wav (\citeauthor{129sotelo2017char2wav}, \citeyear{129sotelo2017char2wav})             & RNN                    & Autoregression                          & GMM   attention                                                                                                 & Low naturalness of synthesized speech                                               \\
		Deep Voice 3 (\citeauthor{6ping2017deep}, \citeyear{6ping2017deep})           & CNN                    & Autoregression                          & Dot-product   attention, positional encoding, attention window                                                  & Attention   is monotonic                                                              \\
		VoiceLoop (\citeauthor{128taigman2017voiceloop}, \citeyear{128taigman2017voiceloop})           & Shifting   buffer      & Autoregression                          & GMM   attention                                                                                                 & Low   naturalness of synthesized speech                                               \\
		DCTTS (\citeauthor{9tachibana2018efficiently}, \citeyear{9tachibana2018efficiently})           & CNN                  & Autoregression                          & Dot-product attention and guided attention                                                                    & Stable, alignment errors are rare                                                   \\
		Tacotron 2 (\citeauthor{5shen2018natural}, \citeyear{5shen2018natural})             & LSTM, CNN            & Autoregression                          & Mixed location-sensitive attention                                                                            & Able to synthesize long sentences accurately                                        \\
		DurIAN (\citeauthor{66yu2019durian}, \citeyear{66yu2019durian})                   & CBHG, RNN            & Autoregression                          & Duration prediction model, external alignment model                                                           & Stable, alignment errors are rare                                                   \\
		FastSpeech (\citeauthor{13ren2019fastspeech}, \citeyear{13ren2019fastspeech})              & Transformer            & Non-autoregression                      & Duration prediction model, knowledge distillation                                                             & Errors will occur when synthesizing long sentences                                  \\
		FastSpeech  2 (\citeauthor{15ren2020fastspeech}, \citeyear{15ren2020fastspeech})            & Transformer            & Non-autoregression                      & Duration prediction model, MFA toolkit                                                                        & Stable, alignment errors are rare                                                   \\
		ParaNet (\citeauthor{120peng2020non}, \citeyear{120peng2020non})                & CNN                    & Non-autoregression                      & Dot-product attention, positional encoding, attention window, multi-layer attention,   knowledge distillation & Attention   alignment is monotonic and stable                                         \\
		EATS (\citeauthor{114donahue2020end}, \citeyear{114donahue2020end})                & CNN                    & GAN                      & Duration prediction model, internal alignment module                                                          & Stable, alignment errors are rare                                                   \\
		Non-Attentive Tacotron (\citeauthor{138shen2020non}, \citeyear{138shen2020non}) & RNN                    & Autoregression                          & Duration prediction model, external alignment module                                                          & Stable, alignment errors are rare                                                   \\
		FastPitch (\citeauthor{112lancucki2020fastpitch}, \citeyear{112lancucki2020fastpitch})          & Transformer            & Non-autoregression                      & Duration prediction model, knowledge distillation                                                             & Can control the pitch contour of synthesized speech                                 \\
		Glow-TTS (\citeauthor{141kim2020glow}, \citeyear{141kim2020glow})               & Transformer, Glow    & Normalizing flow                      & Duration   prediction model,  MAS algorithm                                                                     & The alignment is monotonic and stable                                               \\
		AlignTTS (\citeauthor{135zeng2020aligntts}, \citeyear{135zeng2020aligntts})               & Transformer            & Non-autoregression                      & Duration   prediction model, internal alignment module                                                          & Stable, alignment errors are rare                                                   \\
		SpeedySpeech (\citeauthor{165vainer2020speedyspeech}, \citeyear{165vainer2020speedyspeech})         & CNN                    & Non-autoregression                      & Duration   prediction model, knowledge distillation                                                             & Stable, alignment errors are rare                                                   \\
		JDI-T (\citeauthor{137lim2020jdi}, \citeyear{137lim2020jdi})                   & Transformer            & Non-autoregression                      & Duration   prediction model, knowledge distillation                                                             & Joint training of teacher and student network, stable and alignment errors are rare \\
		TalkNet (\citeauthor{136beliaev2020talknet}, \citeyear{136beliaev2020talknet})             & CNN                    & Non-autoregression                      & Duration   prediction model, ASR model                                                                          & Stable, alignment errors are rare                                                   \\
		Flow-TTS (\citeauthor{179miao2020flow}, \citeyear{179miao2020flow})               & Glow                   & Normalizing flow                      & Multi-head   dot-product attention, internal length predictor                                                   & High   quality of synthesized speech, fast training and inference speed               \\
		DeviceTTS (\citeauthor{192huang2020devicetts}, \citeyear{192huang2020devicetts})             & DFSMN, RNN           & Combination of autoregression and non-autoregression & Duration   prediction model                                                                                     & Stable, alignment errors are rare                                                   \\
		Parallel Tacotron (\citeauthor{207elias2020parallel}, \citeyear{207elias2020parallel})     & LConv                  & Non-autoregression                      & Duration   prediction model, HMM-based aligner                                                                  & Stable, alignment errors are rare                                                   \\
		RobuTrans (\citeauthor{253li2020robutrans}, \citeyear{253li2020robutrans})                & Transformer            & Autoregression                          & Duration   prediction model, speech recognition tools                                                           & Stable, alignment errors are rare                                                  \\
			\noalign{\smallskip}\hline
		\end{tabular}
	\end{spacing}
\end{table*}

\subsection{Expressive acoustic model}
The speech synthesized by deep learning method has a smooth tone, without rhythm and expressiveness, thus it often has a certain gap with the real human voice. In order to synthesize expressive speech, three parts need to be considered: "what to say", "who to say" and "how to say". "What to say" is controlled by the input text and the text front-end. "Who to say" can be controlled by collecting a large amount of voice data of a person and then training the model to learn to imitate the speaker's voice. "How to say" is controlled by prosodic information such as tone, speech rate, and emotion of the synthesized speech. In this paper, "who to say" and "how to say" are collectively referred to as the style features of synthesized speech.
\subsubsection{Acoustic model with reference encoder}
\label{sec:3.3.1}
Style information can be introduced by adding a reference encoder to synthesize expressive speech. There are mainly two methods based on reference encoders that can be used to synthesize speech with a specific style. The first method is to directly control various speech style parameters, such as pitch, loudness, and emotion, by using a trained reference encoder. The second method is to input the reference audio into the reference encoder and use the style parameters encoded by the reference encoder to transfer the speech style features between the reference speech and the target speech. Different methods and models have been proposed to disentangle the different style feature information so that each style feature can be easily controlled individually to synthesize speech with the target style. These methods and models are described in the following paragraphs.\par
\citet{33skerry2018towards} divided the features of speech into three components: text, speaker, and prosody. A reference encoder is added to tacotron to extract the prosody embedding from the reference speech with a specific style, and the speaker embedding is obtained by using a speaker embedding lookup table. Then the prosody embedding, speaker embedding and text embedding are combined and input into the decoder to synthesize speech with the style of the reference speech. \citet{34gururani2019prosody} refined the model on the basis of \citet{33skerry2018towards}, divided the style features of speech into pitch and loudness, and selected two 1-D time series to model the fundamental frequency $f_0$ and loudness of the reference speech respectively. In order to transfer the emotion features in the reference speech more accurately, \citet{184li2021controllable} added two emotion classifiers after the reference encoder and decoder respectively to enhance emotion classification ability in the emotion space. Moreover, they adopted a style loss \cite{185johnson2016perceptual,186gatys2015neural} to measure the style differences between the generated and reference mel-spectrogram \cite{187ma2018neural,188gatys2016image}.\par
Voice conversion (VC) model can disentangle the speaker-dependent timbre feature from speech \cite{117chou2018multi,118chou2019one,116qian2019autovc,123serra2019blow,124kameoka2018stargan}, but cannot extract other style features such as the content, pitch and rhythm of speech. Inspired by the voice conversion model AutoVC \cite{116qian2019autovc}, \citet{119qian2020unsupervised} proposed SPEECHFLOW, which is a speech style conversion model that can disentangle the rhythm, pitch, content, and timbre information. Rhythm, pitch and content features are extracted by three encoders respectively, and timbre feature is represented by one-hot vector of speaker ID. SPEECHFLOW can be trained for speech style conversion by replacing the input of the three encoders with the spectrogram or pitch contour of the reference speech.\par
Similarly, in order to disentangle different style features in speech and achieve the purpose of individually controlling each feature, \citet{35wang2018style} introduced a global style token (GST) network in Tacotron, which plays a role of clustering. When the GST network is trained with speech data with various styles, multiple meaningful and interpretable tokens can be obtained. The weighted sum of these tokens is used as a style embedding to control and transfer the style features of speech. In inference, a specific weight can be chosen directly for each style token, or a reference signal can be fed to guide the choice of token combination weights. For the choice of token weight, \citet{36kwon2019effective} proposed a controlled weight (CW)-based method to define the weight values by investigating the distribution of each emotion in the emotional vector space. \citet{189um2020emotional} proposed to improve the method of simply averaging the style embedding vectors belonging to each emotion category \cite{190kwon2019effective} to determine the representative weight vectors by maximizing the ratio of inter-category distance to intra-category distance (I2I), and proposed to apply the spread-aware I2I (SA-I2I) method to change the emotion intensity instead of the simple linear interpolation-based approach. Mellotron \cite{37valle2020mellotron} additionally introduces fundamental frequency $f_0$ information, and takes text, speaker, fundamental frequency $f_0$, attention mapping, and GST as conditions when synthesizing speech, in which the speaker represents timbre, the fundamental frequency $f_0$ represents pitch, the attention mapping represents rhythm, and GST represents prosody.\par
Since GST-Tacotron uses only paired input text and reference speech for training, inputting unpaired text and speech during synthesis will cause the generated sound to become blurry. Moreover, in this case, the reference encoder may store some text information in the reference embedding rather than prosody and speaker information to reconstruct the input speech. Using the idea of dual learning, \citet{58liu2018improving} proposed to train GST-Tacotron with unpaired text and speech, and input the output mel-spectrogram into the ASR model to predict the input text, thus preventing the reference encoder from encoding any text information. Furthermore, they also use the regularization method of attention consistency loss to accelerate the training convergence speed of both ASR and TTS models.\par
In order to control the style of synthesized speech more flexibly, multiple reference encoders can be used to extract different style features of multiple reference speech respectively. For example, \citet{38bian2019multi} used multiple reference encoders based on GST network to disentangle different style features, and proposed intercross training technique to separate the style latent space by introducing orthogonality constraints between the extracted styles of each encoder. However, this intercross training scheme does not guarantee each combination of style classes is seen during training, causing a missed opportunity to learn disentangled representations of styles and sub-optimal results on disjoint datasets. \citet{39whitehill2019multi} used an adversarial cycle consistency training scheme to ensure the use of information from all style dimensions to address the challenges of multi-reference style transfer on disjoint datasets. They achieved a higher rate of style transfer for disjoint datasets than previous models.\par
Variational auto-encoder (VAE) \cite{51kingma2013auto} generates samples with specific features by sampling from the distribution of latent variables. Latent variables are continuous and can be interpolated, similar to the implicit style features in speech. The speech style features learned by VAE in an unsupervised manner can be easily separated, scaled and combined. Therefore, there are many tasks that use VAE to control the synthesized speech style. The speech style features learned by VAE in an unsupervised manner can be easily separated, scaled and combined. Therefore, there are many works using VAE to control the style of synthesized speech. For example, \citet{40zhang2019learning} added a VAE network to Tacotron 2 to learn latent variables representing speech style. Each dimension of latent variables represents a different style feature. In order to further disentangle the various style features of speech, \citet{41hsu2019disentangling} proposed GMVAE-Tacotron based on the Gaussian mixture VAE network, with two levels of hierarchical latent variables. The first level is a discrete latent variable, representing a certain category of style (e.g. speaker ID, clean/noisy). The second level is a continuous latent variable approximated by the multivariate Gaussian distribution. Each component represents the degree of the feature (e.g. noise level, speaking rate, pitch) under the category of the first level. In general, it is equivalent to using the GMM to fit the distribution of latent variables. This model can effectively factorize and independently control latent attributes underlying the speech signal.\par
However, these methods only model the global style features of speech, without considering prosodic control at the phoneme and word levels. In order to model acoustic features at various resolutions, \citet{194sun2020fully}, in addition to modeling global speech features such as noise and channel number, also modeled word-level and phoneme-level prosodic features such as fundamental frequency $f_0$, energy and duration. They used a conditional VAE with an autoregressive structure to make prosodic features of each layer more interpretable and to impose hierarchical conditioning across all latent dimensions. Parallel Tacotron \cite{207elias2020parallel} used two different VAE models, one similar to \citet{41hsu2019disentangling} for modeling global features of speech such as different prosodic patterns of different speakers, and the other similar to \citet{194sun2020fully} for modeling phoneme-level fine-grained features.\par
Normalizing flow can control the latent variables to synthesize speech with different styles by learning an invertible mapping of data to a latent space. For example, Flowtron \cite{74valle2020flowtron} applied the normalizing flow to Tacotron to control speech variation and style transfer by learning a latent space that stores non-textual information. Glow-TTS \cite{141kim2020glow} takes Glow \cite{22kingma2018glow} as the decoder to control the style of synthesized speech by controlling the prior distribution of latent variables. It is also possible to model speech style features with both normalizing flow and VAE. \citet{42aggarwal2020using} used VAE and Householder Flow \cite{52tomczak2016improving} to improve the reference encoder proposed by \citet{33skerry2018towards}, thereby enhancing the disentanglement capability of the TTS system.\par
GAN can also be used in style speech synthesis. For example, \citet{187ma2018neural} enhanced the content-style disentanglement ability and controllability of the model by combining a pairwise training procedure, an adversarial game, and a collaborative game into one training scheme. The adversarial game concentrates the true data distribution, and the collaborative game minimizes the distance between real samples and generated samples in both the original space and the latent space.\par
\subsubsection{Acoustic model of explicit modeling style features}
The prosody of the speech can also be controlled intuitively by constraining the prosodic features of the waveform. For example, \citet{166morrison2020controllable} proposed a user-controllable, context-aware neural prosody generator that allows the input of the $f_0$ contour for certain time frames and generates the remaining time frames from input text and contextual prosody. CHiVE \cite{139kenter2019chive} is a conditional VAE model with a hierarchical structure. It can generate prosodic features such as fundamental frequency $f_0$, energy $c_0$ and duration suitable for use with a vocoder, and yield a prosodic space from which meaningful prosodic features can be sampled. To efficiently capture the hierarchical nature of the linguistic input (words, syllables and phones), both the encoder and decoder parts of the auto-encoder are hierarchical, in line with the linguistic structure, with layers being clocked dynamically at the respective rates.\par
In practical applications, since it is difficult to interpret and give practical meaning to each of the latent variables learned by unsupervised style separation methods such as GST and VAE, FastSpeech uses a length adjuster to replicate and expand the hidden state of the phoneme sequence according to the duration of each phoneme, thus intuitively controlling the speech speed and some prosodic features.\par
FastPitch \cite{112lancucki2020fastpitch} adds a pitch prediction network to FastSpeech to control pitch. Compared with FastSpeech and FastPitch, FastSpeech 2 introduces more style features such as pitch, energy, and more accurate duration as conditional inputs to construct a variance adaptor, and uses trained predictors of energy, pitch, and duration predictors to synthesize speech with a specific style. Durian simply divides speech styles into several discrete categories, learns embedding vectors from speech data with various styles through supervised learning , and controls the intensity of the style by multiplying a scalar.\par
\subsubsection{Multi-speaker acoustic model}
Multi-speaker speech synthesis is also an important task of TTS model. A simple way to synthesize the voices of multiple speakers is to add a speaker embedding vector to the input \cite{4gibiansky2017deep,6ping2017deep}. The speaker embedding vector can be obtained by additionally training a reference encoder. For example, \citet{43jia2018transfer}, \citet{127arik2018neural} and \citet{130nachmani2018fitting} introduced a speaker encoder in Tacotron 2, Deep Voice 3 and VoiceLoop \cite{128taigman2017voiceloop} respectively to encode the speaker information in the reference speech into a fixed-dimensional speaker embedding vector. The embedding vector can be extracted only from a small number of speech fragments of the target speaker. The speech data corpus used to train the speaker encoder only needs to contain the recordings of a large number of speakers, but does not need to be of high quality. Even if the training data contains a small amount of noise, the extraction of timbre features will not be affected.\par
The speaker adaptation can also be used for multi-speaker speech synthesis. \citet{127arik2018neural}, \citet{128taigman2017voiceloop}, and \citet{156zhang2020adadurian} fine-tune the trained multi-speaker model using a small number of $\langle$text, speech$\rangle$ data pairs of the target speaker. Fine-tuning can be applied to the speaker embedded vector \cite{127arik2018neural,128taigman2017voiceloop}, part of the model \cite{156zhang2020adadurian}, or the whole model \cite{127arik2018neural}. \citet{171moss2020boffin} proposed a fine-tuning method to select different model hyperparameters for different speakers, achieving the goal of synthesizing the voice of a specific speaker with only a small number of speech samples, in which the selection of hyperparameters adopts the Bayesian optimization method \cite{172shahriari2015taking}.\par
However, these methods are not very effective when synthesizing the speech of unseen speakers. To solve this problem, \citet{44cooper2020zero} extracted speaker information by using learnable dictionary encoding (LDE) on the basis of \citet{43jia2018transfer}, and inserted the speaker embedding into both prenet layer and attention network of Tacotron 2 as additional information. When training the speaker encoder, \citet{130nachmani2018fitting} introduced, in addition to the use of MSE losses, the contrast loss term and the cyclic loss term, which allowed the model to synthesize the voice of the new speaker with only a small amount of audio. When training the speaker encoder, in addition to the MSE loss, \citet{130nachmani2018fitting} also a contrastive loss term and a cyclic loss term, which allow the model to synthesize the voice of a new speaker with only a small amount of audio. \citet{216cai2020speaker} and \citet{217shi2020aishell} introduced an identity feedback constraint by adding an additional loss term between the reference embedding and the extraction embedding of the synthesized signal, thus increasing the robustness and speaker similarity of the produced speeches. 
\subsection{Low-resource acoustic model}
Deep learning based acoustic models need to be trained with a large number of high-quality $\langle$text, speech$\rangle$ data pairs to synthesize high-fidelity speech, and the data set requirements are higher when synthesizing speech with specific prosody and emotion. But for $95\%$ of languages and audio with a specific style, the corpus is very scarce. Moreover, the English speech corpus used for TTS usually contains about $10-40$ hours of speech data and contains no more than $20,000$ words. The largest public English speech corpus, LibriTTS \cite{221zen2019libritts}, contains only $80,000$ words, which is far lower than the number of words in the regular English vocabulary (usually $130,000-160,000$). When synthesizing, the acoustic model may mispronounce words outside the training set. It is difficult to cover all vocabulary just by increasing the number of training utterances, because the natural frequency of words tends to follow the Zipfian distribution \cite{28taylor2019analysis}, which means that the number of new words contained in the speech data per hour gradually decreases. Therefore, to achieve a linear increase in word coverage would require an exponential increase in audio data, which would be costly and impractical. Besides, most speech data is recorded by non-professionals and contains a lot of noise. Therefore, the lack of high-quality speech training data in TTS is mainly manifested in the lack of training data that cannot cover all vocabulary and contains noise.\par
To solve the problem that the speech data cannot cover all the words, text and phonemes can be input into the acoustic network together. During training, some words can be represented by text randomly, so that the acoustic model can predict the phoneme pronunciation of unseen words according to the learned correspondence between characters and phonemes \cite{6ping2017deep,120peng2020non}. The text front-end can also be used to convert the text into phonemes in advance, in order to make the model only need to learn the pronunciation of a small number of phonemes.\par
To solve the problem of the lack of speech data for minority languages and dialects, the method of cross-language transfer learning can be used. For example, \citet{46guo2018dnn} and \citet{47zhang2019deep} trained an average language model with a large Mandarin corpus and a small Tibetan corpus when training the Tibetan TTS model, which made up for the lack of Tibetan speech data. \citet{72tu2019end} introduced cross-language transfer learning into Tacotron. They used speech data from high-resource languages to pre-train Tacotron, and then fine-tuned the pre-trained model with speech data from low-resource languages. \citet{167nekvinda2020one} used the idea of meta-learning to train the acoustic model with only a small number of samples from multiple languages in order to synthesize speech containing multiple languages. They used a fully convolutional encoder from DCTTS, whose parameters are generated using a separate contextual parameter generator network \cite{168platanios2018contextual} conditioned on language embedding, thus realizing cross-lingual knowledge-sharing.\par
Semi-supervised pre-training can also be used to reduce the demand of the TTS model for paired training data. \citet{67chung2019semi} proposed training the encoder and decoder with unpaired text and speech respectively, and then fine-tuning the pre-trained Tacotron with a small amount of $\langle$text, speech$\rangle$ data pairs. Although this approach helps the model synthesizes more intelligible speech, the experimental results show that pre-training the encoder and decoder separately at the same time does not bring further improvement than only pre-training the decoder. And there is a mismatch between pre-training only the decoder and fine-tuning the whole model, because during pre-training the decoder is only conditioned on the previous frame, while during fine-tuning the decoder is also conditioned on the text representation output by the encoder. To avoid potential error caused by this mismatch and further improve the data efficiency by using only speech, \citet{69zhang2020unsupervised} proposed to use vector Vector-quantization Variational-Autoencoder (VQ-VAE) \cite{70chorowski2019unsupervised,71oord2017neural} to extract unsupervised linguistic units from untranscribed speech and then use $\langle$linguistic units, speech$\rangle$ pairs to pre-train the entire model. The language units act as phonemes that are paired with the audio, while VQ-VAE plays a role similar to speech recognition model. However, VQ-VAE is trained in an unsupervised way to obtain discretized linguistic representations, which is suitable for low-resource languages. Finally, the model is fine-tuned with a small amount of $\langle$text, speech$\rangle$ data pairs.\par
Using dual learning to train TTS and ASR models simultaneously can also achieve the purpose of using text or speech data alone to train both models. \citet{57tjandra2017listening} proposed an auto-encoder model in which one is regarded as an encoder and the other as a decoder. For example, when there is only speech but no corresponding text, the ASR model can be used as the encoder to output text, and the TTS model can be used as the decoder to output speech, and then the speech output of the TTS model is expected to be as close as possible to the input speech. The other situation is similar when there is only text but no speech. \citet{68ren2019almost} also used the idea of dual learning to combine TTS and ASR to build the capability of the language understanding and modeling in both speech and text domains using unpaired data during training, that is, using denoising auto-encoder (DAE) to reconstruct corrupt speech and text in an encoder-decoder framework. They also used a dual transformation (DT) approach similar to \citeauthor{57tjandra2017listening} to train the model to convert text to speech and speech to text respectively. The difference is that \citeauthor{57tjandra2017listening} relied on two well-trained TTS and ASR models, whereas \citeauthor{68ren2019almost} trained the two models from scratch, which is suitable for the lack of training data.\par
Multi-speaker TTS model has much lower requirements on the quantity and quality of training data than models that synthesize speech with a specific style, because they only need to separate and capture timbre information in the audio. However, if the speech training data of the target speaker is too small, the timbre features cannot be effectively learned. In order to increase the amount of speech data of the target speaker, \citet{254huybrechts2020low} used a voice conversion (VC) model to convert the voice data of other speakers into the voice of the target speaker for data augmentation, then trained the TTS model with the expanded speech data, and finally used the real voice data of the target speaker for fine-tuning.\par
The noise in the training data can be reduced by pre-processing steps. \citet{48valentini2018speech} took the acoustic features of clean speech and noisy speech respectively as the input and target of RNN network, enabling the network to convert noisy speech into clean speech. Generally, the data in the corpus containing different styles of speech is of low quality and contains noise, which will hinder the training of style speech synthesis model. In this case, the method of speech style control introduced in Sect. \ref{sec:3.3.1} can be used to train the style extraction network with clean data and noisy data. By making the network learn about the latent variables of noise features, it can synthesize clean speech. For example, \citet{35wang2018style} trained GST-Tacotron by using data sets mixed with various noises to learn tokens about the noise features. During synthesis, the token representing noiseless is used as the style embedding to convert noisy reference speech to clean speech. \citet{41hsu2019disentangling} used one dimension of the mixed Gaussian distribution to represent the noise feature. Clean speech can be synthesized by using the average value of the clean speech class or the value of the noise variable extracted from the clean reference speech as the value of the noise feature.
\section{Vocoder}
\label{sec:4}
Inspired by the successful application of autoregressive generative model \cite{49van2016pixel} in the field of image and natural language generation, \citet{17oord2016wavenet} first applied this method to TTS and proposed the most widely used vocoder WaveNet. In order to capture the long-range temporal dependencies in audio signals, WaveNet adopts a multi-layer dilated causal gated convolutional network, which makes the receptive field to grow exponentially with the depth. WaveNet uses speaker identity and linguistic features as global and local conditions respectively to synthesize the speech of the target speaker and text. However, WaveNet has a complex network structure and is autoregressive, therefore the training and inference speed is slow. Moreover, the speech synthesized with WaveNet is sometimes not natural. Therefore, after it was proposed, there are a lot of work to improve it. The direction of improvement is mainly to accelerate the speed of training and inference and improve the quality of synthesized speech, which are respectively called fast vocoder and high-quality vocoder. These methods are introduced in the following sections.
\subsection{Fast vocoder}
The training can be accelerated by reducing the size and parameters of the vocoder, and the inference can be accelerated by replacing the autoregressive method in WaveNet with non-autoregressive methods. The following sections will introduce various small-size vocoders and non-autoregressive vocoders.
\subsubsection{Small size vocoder}
To improve the speed of training and inference, Fftnet \cite{53jin2018fftnet} uses the simple ReLU activation function and $1\times1$ convolutions to replace the gated activation units and dilated convolutions in WaveNet, which reduces the computational cost. SampleRNN \cite{18mehri2016samplernn} adopted a multi-scale RNN structure. Different layers operate on audio data of different time scales. Compared with WaveNet, it only processes individual samples in the last layer to improve the synthesis speed, and back-propagates the gradient of the loss function only on a small fraction of audio to improve the training speed. WaveRNN \cite{269kalchbrenner2018efficient} only uses a single-layer GRU network with a dual softmax layer that predict respectively the 8 coarse (or more significant) bits and the 8 fine (or least significant) bits of the 16-bit audio sample, and applies a weight pruning technique to further reduce the model parameters. Furthermore, for the purpose of generating multiple channels of speech in parallel, WaveRNN divides a long audio sequence into multiple short sequences evenly during inference, and the generation within and between each short sequence is autoregressive. Although WaveRNN is autoregressive and based on RNN, its training and inference time is still short, thus it can be used in systems with few resources such as mobile phones and embedded systems. The Multi-Band WaveRNN proposed by \citet{66yu2019durian} further improves the inference speed of WaveRNN by generating multiple bands in parallel, and performs 8-bit quantization on the weight value to reduce the model size. LPCNet \cite{157valin2019lpcnet} reduces the complexity of the model by combining WaveRNN with linear prediction (LP) technology in traditional digital signal processing, thereby improving the synthesis efficiency. The characteristics of various small-size vocoders are shown in Table \ref{tab:3}.

\begin{table*}
	\small
	\caption{Small size vocoder}
	\label{tab:3}       % Give a unique label
	\begin{spacing}{1.4}
		\begin{tabular}{|p{3cm}|p{2.5cm}|p{6cm}|}
			\hline\noalign{\smallskip}
		Vocoder                                & Neural   network types     & Characteristics                                                                                                                \\
			\noalign{\smallskip}\hline\noalign{\smallskip}
		WaveNet (\citeauthor{17oord2016wavenet}, \citeyear{17oord2016wavenet})          & Dilated   causal gated CNN & Based   on dilated CNN, the training and inference speed is slow                                                               \\
		SampleRNN   (\citeauthor{18mehri2016samplernn}, \citeyear{18mehri2016samplernn})       & RNN                        & Multi-scale   RNN structure, training and inference speed is faster than Wavenet                                               \\
		FftNet   (\citeauthor{53jin2018fftnet}, \citeyear{53jin2018fftnet})            & $1\times1$ CNN                    & Based   on $1\times1$ convolution, the model structure is simple, and the training and inference   speed is fast                      \\
		WaveRNN (\citeauthor{269kalchbrenner2018efficient}, \citeyear{269kalchbrenner2018efficient})   & GRU                        & Based   on single layer of GRU, the model structure is simple, and the training and inference   speed is fast                  \\
		Multi-Band   WaveRNN (\citeauthor{66yu2019durian}, \citeyear{66yu2019durian}) & GRU                        & Parallel   generation of multiple bands, the training and inference speed is fast                                              \\
		LPCNet   (\citeauthor{157valin2019lpcnet}, \citeyear{157valin2019lpcnet})          & GRU                        & The   linear prediction (LP) technology is used, the model structure is simple, and   the training and inference speed is fast\\
			\noalign{\smallskip}\hline
		\end{tabular}
	\end{spacing}
\end{table*}

\subsubsection{Non-autoregressive vocoder}
\label{sec:4.1.2}
Similar to acoustic models, these vocoders increase the speed of training and inference to a certain extent, but all generate audio signals frame by frame in an autoregressive manner. If the non-autoregressive generation method can be used to generate speech waveforms in parallel, the inference speed will be greatly improved. Based on this idea, various non-autoregressive vocoders are proposed, and their characteristics are shown in Table \ref{tab:4}.
\par
\begin{table*}
	\small
	\caption{Non-autoregressive vocoder}
	\label{tab:4}       % Give a unique label
	\begin{spacing}{1.5}
		\begin{tabular}{|p{2.5cm}|p{3cm}|p{2cm}|p{6cm}|}
			\hline\noalign{\smallskip}
		Vocoder                                    & Neural   network types                                             & Generative   model types      & Characteristics                                                                                                                                                           \\ \noalign{\smallskip}\hline\noalign{\smallskip}
		WaveNet   (\citeauthor{17oord2016wavenet}, \citeyear{17oord2016wavenet})               & Dilated   causal gated convolution                                 & Autoregression                & Autoregressive   generation, slow training and inference speed                                                                                                            \\
Parallel   WaveNet (\citeauthor{20oord2018parallel}, \citeyear{20oord2018parallel})     & Dilated   causal gated convolution                                 & IAF                           & Based   on knowledge distillation, training and inference speed is fast, Monte Carlo   sampling is required to estimate KL divergence, the training process is   unstable \\
FloWaveNet   (\citeauthor{125kim2018flowavenet}, \citeyear{125kim2018flowavenet})            & Dilated   convolution                                              & Normalizing   flow            & The   inference speed is fast, the training convergence speed is slow, the model contains   many parameters                                                               \\
ClariNet   (\citeauthor{25ping2018clarinet}, \citeyear{25ping2018clarinet})             & Dilated   causal gated convolution                                 & IAF                           & Based   on knowledge distillation, the training and inference speed is fast, the   training process is stable                                                             \\
WaveGlow   (\citeauthor{21prenger2019waveglow}, \citeyear{21prenger2019waveglow})          & Non-causal   dilated convolution, $1\times1$ convolution                  & Normalizing   flow            & The   inference speed is fast, the training convergence speed is slow, the model contains   many parameters                                                               \\
MelGAN   (\citeauthor{24kumar2019melgan}, \citeyear{24kumar2019melgan})              & Dilated   convolution, transposed convolution, grouped convolution & GAN                           & The   inference speed is fast, the training convergence speed is slow                                                                                                     \\
GAN-TTS   (\citeauthor{113binkowski2019high}, \citeyear{113binkowski2019high})         & Dilated   convolution                                              & GAN                           & The   training and inference speed is fast, no need for mel-spectrogram as input                                                                                          \\
Parallel   WaveGAN (\citeauthor{23yamamoto2020parallel}, \citeyear{23yamamoto2020parallel}) & Non-causal   dilated convolution                                   & GAN                           & The   inference speed is fast, the training convergence speed is slow, the model contains   many parameters                                                               \\
WaveVAE   (\citeauthor{120peng2020non}, \citeyear{120peng2020non})              & Dilated   causal gated convolution                                 & IAF,   VAE                    & The   training and inference speed is fast                                                                                                                                \\
WaveFlow   (\citeauthor{121ping2020waveflow}, \citeyear{121ping2020waveflow})             & 2D-dilated convolution                                           & Autoregression                & Combining   the advantages of autoregressive flow and non-autoregressive flow, the   training and inference speed is fast                                                 \\
WaveGrad   (\citeauthor{160chen2020wavegrad}, \citeyear{160chen2020wavegrad})             & Dilated   convolution                                              & Diffusion   probability model & The   inference speed is fast, the training convergence speed is slow                                                                                                     \\
DiffWave   (\citeauthor{244kong2020diffwave}, \citeyear{244kong2020diffwave})             & Bidirectional   dilated convolution                                & Diffusion   probability model & The   inference speed is fast, the training convergence speed is slow                                                                                                     \\
Multi-Band   MelGAN (\citeauthor{180yang2021multi}, \citeyear{180yang2021multi})    & Dilated   convolution, transposed convolution, grouped convolution & GAN                           & The   training and inference speed is fast                                                                                                                               \\
			\noalign{\smallskip}\hline
		\end{tabular}
	\end{spacing}
\end{table*}

The traditional Gaussian autoregressive model is equivalent to an autoregressive flow (AF) \cite{62kingma2016improving}, which is a kind of normalizing flow \cite{65rezende2015variational}. The main idea of the normalizing flow is that a complex distribution can be obtained by a simple distribution transformed through multiple invertible functions. It was originally proposed to make the distribution function of latent variables in VAE \cite{51kingma2013auto} more complex. The flow-based generative model learns the bidirectional mapping from the input sample $x$ to the latent representation $\boldsymbol{z}$, i.e. $\boldsymbol{x}=f(\boldsymbol{z})$ and $\boldsymbol{z}=\boldsymbol{f}^{-1}(\boldsymbol{x})$. This mapping $\boldsymbol{f}$ is called a normalizing flow and is an invertible function fitted by neural networks, consisting of $k$ invertible transformations $\boldsymbol{f}=f_1\circ \cdots \circ f_k$. The normalizing flow transforms a simple density distribution $p(\boldsymbol{z})$ (such as an isotropic Gaussian distribution) to a complex distribution $p(\boldsymbol{x})$ by applying an invertible transformation $\boldsymbol{x}=\boldsymbol{f}(\boldsymbol{z})$. The probability density of $\boldsymbol{x}$ can be calculated through the change of variables formula:
\begin{equation}
	p(\boldsymbol{x})=p(\boldsymbol{z})\Big\vert det\Big(\frac{\partial \boldsymbol{f}^{-1}(\boldsymbol{x})}{\partial \boldsymbol{x}}\Big) \Big\vert
\end{equation}
where $det$ is the Jacobian determinant. The computation of the determinant has the complexity of $O(n^3)$, where $n$ is the dimension of $\boldsymbol{x}$ and $\boldsymbol{z}$. In order to reduce the amount of computation, two flow models that can easily calculate the Jacobian determinant have been proposed, respectively, based on autoregressive transformation \cite{49van2016pixel} and bipartite transformation \cite{22kingma2018glow,63dinh2014nice,64dinh2016density}.\par
During training, the autoregressive flow calculates the latent variable $z_i,i=1,\ldots,D$ by transforming the speech $\boldsymbol{x}={x_1,x_2,\ldots,x_D}$:
\begin{equation}
	z_i=\sigma_i(x_{1:i-1})\cdot x_i+\mu_i(x_{1:i-1})
\end{equation}
where $z_{1:D}$ is $D$ latent variables subject to the isotropic Gaussian distribution, $\mu$ is the shift variables representing the mean, and $\sigma$ is the scaling variables representing the standard deviation. The training process is non-autoregressive, and $z_i$ only depends on $x_{1:i}$. In this case, the Jacobian matrix is a triangular matrix whose determinant is the product of the diagonal terms:
\begin{equation}
	det\Big(\frac{\partial f^{-1}(\boldsymbol{x})}{\partial \boldsymbol{x}}\Big)=\prod_i\sigma_i(x_{1:i-1})
\end{equation}
During inference, the trained $z_i,i=1,\ldots,D$ and the previously generated audio $x_{1:i-1}$ are used to predict the new $x_i$:
\begin{equation}
	x_i=\frac{z_i-\mu_i(x_{1:i-1})}{\sigma_i(x_{1:i-1})}
\end{equation}
This inference process is autoregressive, resulting in slow inference. In order to speed up the inference speed, Parallel WaveNet \cite{20oord2018parallel} and its improved model ClariNet \cite{25ping2018clarinet} use inverse autoregressive flow (IAF) \cite{62kingma2016improving} to generate speech in Parallel. IAF is another normalizing flow. In contrast to AF, IAF uses the previously obtained latent variable $z_{1:i-1}$ to calculate $z_i$ during training:
\begin{equation}
	z_i=\frac{x_i-\mu_i(z_{1:i-1})}{\sigma_i(z_{1:i-1})}
\end{equation}
This training process is autoregressive. In inference, $z_{1:i}$ is used to predict $x_i$:
\begin{equation}
	x_i=\sigma_i(z_{1:i-1})\cdot z_i+\mu_i(z_{1:i-1})
\end{equation}
This inference process is non-autoregressive. Therefore, AF is fast in training and slow in inference, whereas IAF is just the opposite. In order to train and synthesize quickly at the same time, Parallel WaveNet and ClariNet take the autoregression WaveNet as the teacher network, which is responsible for providing the guidance information on the distribution of $z_i,i=1,\ldots,D$ during training. And IAF is used as the student network to take charge of the final audio sampling, and solve the problem that IAF cannot be trained in parallel by means of probability density distillation.\par
However, due to the knowledge distillation used in Parallel WaveNet and Clarinet, the training process is complex. In order to simplify the training process, \citet{120peng2020non} proposed WaveVAE. The encoder and decoder of WaveVAE are respectively parameterized by a Gaussian autoregressive WaveNet and the one-step-ahead predictions from an IAF. It can jointly optimize the encoder $q_\varphi(\boldsymbol{z}|\boldsymbol{x})$ and decoder $p_\theta(\boldsymbol{x}|\boldsymbol{z})$ to be trained from scratch by maximizing the evidence lower bound (ELBO) for observed $\boldsymbol{x}$ in VAE, but at the expense of sound quality.\par
In order to train and synthesize more quickly, \citet{121ping2020waveflow} proposed WaveFlow which combines autoregressive flow and non-autoregressive convolution. The training process does not need complex knowledge distillation, only based on the likelihood function, and combines the advantages of autoregressive and non-autoregressive flow. It can train and synthesizing high- fidelity speech quickly, while only occupying small memory. WaveFlow represents a 1-D audio sequence $\boldsymbol{x}={x_1,x_2,\ldots,x_D}$ with a 2-D matrix $\boldsymbol{X}\in \boldsymbol{R}^{h\times w}$, in which adjacent samples are in the same column. The latent variable matrix $\boldsymbol{Z}\in \boldsymbol{R}^{h\times w}$ is defined as:
\begin{equation}
	Z_{i,j}=\sigma_{i,j}(X_{1:i-1,:})\cdot X_{i,j}+\mu_{i,j}(X_{1:i-1,:})
\end{equation}
where $X_{1:i-1,:}$ represents all the elements above the $i$-th row. Therefore, the value of $Z_{i,j}$ depends only on the sample in $i$-th row and $j$-th column and the samples above $i$-th row, which can be calculated at the same time. In inference, the sample is generated by:
\begin{equation}
	X_{i,j}=\frac{Z_{i,j}-\mu_{i,j}(X_{1:i-1,:})}{\sigma_{i,j}(X_{1:i-1,:})}
\end{equation}
Although it is autoregressive, it only takes $h$ steps to generate all samples, and $h$ is usually small, like 8 or 16. WaveFlow uses a 2-D dilated CNN to model a 2-D matrix. Non-causal CNN is used on width dimension, causal CNN with autoregressive constraints is used on height dimension, and convolution queue \cite{122paine2016fast} is used to cache the intermediate hidden states to speed up the autoregressive synthesis on height dimension. Therefore, it not only retains both the advantage of autoregressive inference method that can accurately simulate the local variations of waveform and non-autoregressive convolutional structure that can do speedy synthesis and capture the long-range structure in the data.\par
WaveGlow \cite{21prenger2019waveglow} and FloWaveNet \cite{125kim2018flowavenet} are also based on normalizing flow and have similar structures, using Glow \cite{22kingma2018glow} and Real-NVP \cite{64dinh2016density} respectively. Real-NVP is an improved model of the normalizing flow NICE \cite{63dinh2014nice}. It is trained and inferred by bipartite transformation, but each layer can only transform a part of the input. As an improved model of Real-NVP, Glow introduces $1\times1$ invertible CNN to mix the information between two channels and realizes complete transformation. The affine coupling layer in WaveGlow and Flowavenet transforms one half dimension $x_b$ of input vector $\boldsymbol{x}$ each time, leaving the other half dimension $x_a$ unchanged. The transformation process is:
\begin{equation}
	z_a=x_a
\end{equation}
\begin{equation}
	z_b=x_b\cdot \sigma_b(x_a)+\mu_b(x_a)
\end{equation}
where $x_a$ and $x_b$ are the result of bisecting $\boldsymbol{x}$, $z_a$ and $z_b$ are the corresponding latent variables respectively. The inference process is:
\begin{equation}
	x_a=z_a
\end{equation}
\begin{equation}
	x_b=\frac{z_b-\mu_b(x_a)}{\sigma_b(x_a)}
\end{equation}
Therefore, WaveGlow and FloWaveNet can both compute the latent variable $\boldsymbol{z}$ and synthesize the speech frame $\boldsymbol{x}$ in parallel. In fact, the bipartite transformation is a special autoregressive transformation \cite{121ping2020waveflow}, which can be reduced to a bipartite transformation by substitution:
\begin{equation}
	\left( \begin{array}{l}
	\mu_i(x_{1:i-1}) \\
	 \sigma_i(x_{1:i-1})
\end{array} \right)
	=\left\{ \begin{array}{ll}
		(0,1)^T, & i\in a\\
		(\mu_i(x_a),\sigma_i(x_a))^T, & i\in b
	\end{array} \right.
\end{equation}
Compared with autoregressive transformation, bipartite transformation is not as expressive as autoregressive transformation, because it reduces the dependence between data $\boldsymbol{X}$ and latent variable $\boldsymbol{Z}$. As a result, the speech synthesized by WaveGlow and FloWaveNet is of low quality. A deeper network is needed to obtain the results comparable to the autoregressive model.\par
In addition to normalizing flow, GAN \cite{132goodfellow2014generative} can also be used to synthesize speech in parallel, such as Parallel WaveGAN \cite{23yamamoto2020parallel}, MelGAN \cite{24kumar2019melgan}, multi-band MelGAN \cite{180yang2021multi} and GAN-TTS \cite{113binkowski2019high}. Parallel WaveGAN's generator is similar in structure to WaveNet, which uses random noise and mel-spectrogram conditions to generate speech waveforms. Its discriminator is used to determine whether the generated audio is real. MelGAN's generator simply uses dilated CNN to increase the receptive field, and its inference speed is faster than Parallel WaveGAN. Its discriminator outputs real/fake labels and feature maps \cite{183wang2018high}, and speeds up training by using grouped convolutions to reduce the model parameter.\par 
The feature matching loss adopted by MelGAN generates feature maps with neural networks, while the multi-resolution STFT loss adopted by Parallel WaveGAN uses STFT algorithm to generate feature maps. Inspired by this, multi-band MelGAN introduces the multi-resolution STFT loss in Parallel Wavegan into MelGAN instead of the original feature matching loss, and carries out a multi-band extension to MelGAN to measure the difference between the real and predicted audio in multiple subband scales of audio, which further improves the training and inference speed of MelGAN. In order to obtain better results and faster training speed, GAN-TTS uses an ensemble of small scale unconditional and conditional Random Window Discriminators (RWDs) operating at different window sizes, which respectively assess the realism of the generated speech and its correspondence with the input text.\par
The diffusion probability model \cite{163sohl2015deep,164ho2020denoising} can also be used to generate speech waveforms. It is a probabilistic model based on Markov chain, which divides the mapping relationship between the noise and the target waveform into several steps, and gradually transforms the simple distribution (e.g., isotropic Gaussian) into the complex data distribution by means of Markov chain. It first trains the diffusion process of Markov chains (from structured waveform to noise), and then decodes the noise through the reverse process (from noise to structured waveform). The decoding process requires only a constant few generation steps, so the inference speed is fast. \citet{160chen2020wavegrad} proposed a fully convolutional vocoder WaveGrad to synthesize speech non-autoregressively based on diffusion probability model and score matching framework \cite{161song2020sliced,162song2020improved}. A similar model is DiffWave \cite{244kong2020diffwave}, which uses bidirectional dilated convolution architecture with a long bidirectional receptive field and a much smaller number of model parameters than WaveGrad. However, the inference speed of the vocoder based on diffusion probability model is slightly lower than that of the flow-based vocoder. 
\subsection{High-quality vocoder}
To improve the naturalness of speech, WaveNet proposes to expand the receptive field by dilated CNN and introduce additional conditional information, such as speaker information (global conditioning) and acoustic features (local conditioning), by modeling the conditional probability of audio. WaveNet takes softmax layer as the output layer of the network, and adopts nonlinear quantization method of $\mu$-law companding transformation to obtain discrete-value speech signals. Although the reconstructed speech signal is close to the original, the quantization process still introduces white noise into the original signal. \citet{31yoshimura2018mel} proposed a quantization noise shaping method based on mel-cepstrum, which solved this problem by preprocessing WaveNet with a mel-log spectral approximation (MLSA) filter \cite{93imai1983mel}. Because the mel-cepstrum matches the human auditory perception characteristics, this method effectively filters the white noise introduced by the commonly used quantization method in the speech waveform synthesis system, and has no extra computational cost compared with WaveNet in the synthesis stage.\par
In order to improve the quality of the speech synthesized by the autoregressive vocoder, \citet{53jin2018fftnet} proposed to add zero padding to the input to make the network have a stronger generalization ability. And when outputting the result, instead of directly taking the value of the maximum probability, sampling is conducted according to the probability distribution to simulate the real speech signal containing noise. Due to the training error of vocoder, there is always noise in the generated speech sample. And in the process of autoregressive generation, the noise in the synthesized speech sample will become more and more loud over time. Generating new samples with noisy speech samples as input to the network adds more and more uncertainty to the network. Therefore, during the training, they added some noise to the input to make the network robust to the input samples containing noise, and reduced the noise injected into the pronunciation samples by post-processing with spectral subtraction noise reduction \cite{19loizou2013speech}.\par
When using implicit generative models such as GAN to generate audio, speech waveforms of different resolutions can be predicted at the same time to perfect the details of synthesized speech and stabilize the training process, as shown in Table \ref{tab:5}. Parallel Wave GAN and Multi-Band MelGAN use a multi-resolution STFT loss for training. The discriminator in MelGAN adopts a multi-scale structure to simultaneously discriminate feature maps of audio waveforms with different sampling frequencies to learn the features of different audio frequency ranges. Besides, MelGAN uses feature matching loss to optimize both discriminator and generator, thereby reducing the distance between the feature maps of the real and synthesized audio. VocGAN \cite{181yang2020vocgan} uses both multi-resolution STFT loss and feature matching loss, and extends the generator on the basis of MelGAN to output multiple waveforms of different scales. It helps the generator learn the mapping of both low- and high-frequency components of acoustic features by training the generator with the adversarial loss calculated by a set of discriminators with different resolutions. Moreover, VocGAN also applied the joint conditional and unconditional (JCU) loss \cite{182zhang2018stackgan++}. The conditional loss leads the generator to map the acoustic feature of the input mel-spectrogram to the waveform more accurately, thus reducing the discrepancy between the acoustic characteristics of the input mel-spectrogram and the output waveform. In addition to using the multi-scale discriminator in MelGAN, HiFi-GAN \cite{191kong2020hifi} introduced the multi-period discriminator (MPD) to model the periodic patterns of speech. Each sub-discriminator only accepts equally spaced samples of an input audio, aiming to capture different implicit structures from each other by looking at different parts of the input audio. Besides, the generator in HiFi-GAN is connected with a multi-receptive field fusion (MRF) module after each transposed convolution, which can observe patterns of various lengths in parallel. \citet{209gritsenko2020spectral} proposed a method for training parallel vocoder based on the spectral generalized energy distance (GED) \cite{210sejdinovic2013equivalence,211shen2020exact,212gneiting2007strictly} between the generated and the real audio distribution. The main difference from other spectrogram-based losses is that, in addition to the attractive term between the generated data and the actual data, GED also adds a repulsive term between generated data to the training loss to avoid generated samples collapsing to a single point, thus capturing the full data distribution. GED can be combined with the adversarial loss to further improve the synthesized speech quality.\par
\begin{table*}
	\small
	\caption{Methods of GAN-based vocoder to improve the naturalness of generated speech}
	\label{tab:5}       % Give a unique label
	\begin{spacing}{1.4}
		\begin{tabular}{|p{3.5cm}|p{8cm}|}
			\hline\noalign{\smallskip}
		Vocoder                                    & Characteristics                                                                                           \\ \noalign{\smallskip}\hline\noalign{\smallskip}
		MelGAN   (\citeauthor{24kumar2019melgan}, \citeyear{24kumar2019melgan})              & Using   multi-scale discriminant structure and feature matching loss                                      \\
		Parallel   WaveGAN (\citeauthor{23yamamoto2020parallel}, \citeyear{23yamamoto2020parallel})  & Using   multi-resolution STFT loss                                                                        \\
		VocGAN   (\citeauthor{181yang2020vocgan}, \citeyear{181yang2020vocgan})               & Using   multi-resolution STFT loss, feature matching loss, multi-scale waveform   generator, and JCU loss \\
		HiFi-GAN   (\citeauthor{191kong2020hifi}, \citeyear{191kong2020hifi})             & Using   multi-scale discrimination, multi-period discrimination, and MRF                                  \\
		Multi-Band   MelGAN (\citeauthor{180yang2021multi}, \citeyear{180yang2021multi})    & Using   multi-resolution STFT loss                                                                        \\
			\noalign{\smallskip}\hline
		\end{tabular}
	\end{spacing}
\end{table*}

Similar to the acoustic models, the multi-speaker TTS task can also be performed only by the vocoder. \citet{131chen2018sample} borrowed the idea of meta-learning and proposed three methods to synthesize the voice of a new speaker using only a small amount of the target speaker's speech. The first method is to fix other parameters of the model and update only the speaker embedding vector. The second method is to fine-tune all the parameters of the model. The third method is to use a trained neural network encoder to predict the speaker embedding. The experimental results show that the speech synthesized by the second method has the highest naturalness. However, the method they proposed only works when the quality of the training speech data is high.
\section{Speech corpus}
\label{sec:5}
The proposal of the end-to-end TTS method based on deep learning reduces the difficulty of developing a high-quality TTS system. Compared with the ASR model, TTS model requires more high-quality speech data with labels to achieve better training results, and the number of open source corpora that meets these conditions is very small. For the convenience of researchers to carry out experiments, several commonly used open source TTS corpora are introduced below. The details of each corpus are shown in Table \ref{tab:6}.
\begin{table*}
	\small
	\caption{Details of each corpus}
	\label{tab:6}       % Give a unique label
	\begin{spacing}{1.4}
		\begin{tabular}{|p{2cm}|p{2cm}|p{1.5cm}|p{1cm}|p{3cm}|p{1.5cm}|}
			\hline\noalign{\smallskip}
					Corpus       & Language                                                                                & Number of speakers          & Hours & Labeling method                                                & Sampling
					Rate (kHz) 
					 \\ \noalign{\smallskip}\hline\noalign{\smallskip}
		VCTK         & English                                                                                 & 109                         & 44    & Characters                                                     & 48     \\
LJ Speech    & English                                                                                 & 1                           & 24    & Original and standardized characters and   phonemes            & 22.05  \\
LibriTTS     & English                                                                                 & 2,456                        & 585   & Original and standardized characters,   contextual information & 24     \\
CMU ARCTIC   & English                                                                                 & 7                           & 7     & Characters                                                     & 16     \\
Blizzard2011 & English                                                                                 & 1                           & 16.6  & Characters                                                     & 16     \\
Blizzard2013 & English                                                                                 & 1                           & 300   & Characters                                                     & 44.1   \\
Blizzard2017 & English                                                                                 & 1                           & 6     & Characters                                                     & 44.1   \\
CSMSC        & Mandarin                                                                                & 1                           & 12    & Pinyin, rhythm and phoneme boundary                           & 48     \\
AISHELL-3    & Mandarin                                                                                & 218                         & 85    & Characters, pinyin                                             & 44.1   \\
DiDiSpeech   & Mandarin                                                                                & 6,000                        & 800   & Standardized Pinyin                                            & 48     \\
CSS10        & German, Greek, Spanish, French, Chinese,   Japanese, Russian, Finnish, Hungarian, Dutch & Single speaker per language &       & Original and standardized characters                           & 22     \\
Common Voice & 60 languages                                                                            &                             & 7,335  & Characters                                                     & 48    \\
			\noalign{\smallskip}\hline
		\end{tabular}
	\end{spacing}
\end{table*}
\subsection{English speech corpus}
Due to the versatility of English, the academic research on English TTS is the most. Therefore, there are many English TTS corpora available for free, such as VCTK \cite{218veaux2016superseded}, LJ Speech \cite{219ljspeech17} and LibriTTS \cite{221zen2019libritts}.\par
The VCTK corpus\footnote{The VCTK corpus can be freely available for download from \url{https://datashare.is.ed.ac.uk/handle/10283/2119}.} includes speech data uttered by 109 native speakers of English with various accents. Each speaker reads out about 400 sentences, most of which were selected from a newspaper plus the Rainbow Passage and an elicitation paragraph intended to identify the speaker's accent. The speaker uses an omni-directional head-mounted microphone to record speech in a hemi-anechoic chamber of the University of Edinburgh at a sampling frequency of 24 bit and 96 kHz. All recordings were converted into 16 bit, downsampled to 48 kHz, and manually end-pointed. The VCTK corpus was originally recorded for building HMM-based multi-speaker TTS systems.\par
LJ Speech\footnote{The LJ Speech corpus is freely available for download from \url{https://keithito.com/LJ-Speech-Dataset/}.} is a public domain corpus consisting of 13,100 short audio clips of a single speaker, made up of non-professional audiobooks from the LibriVox project \cite{220kearns2014librivox}. Each audio file is a single-channel 16 bit PCM WAV with a sampling rate of 22,050 Hz. The audio clips range in length from approximately 1 second to 10 seconds and are segmented automatically based on silences in the recording, with a total duration of about 24 hours. Clip boundaries generally align with sentence or clause boundaries. The text was matched to the audio manually, and a QA pass was done to ensure that the text accurately matched the words spoken in the audio. \par
The LibriTTS corpus\footnote{The LibriTTS corpus is freely available for download from \url{http://www.openslr.org/60/}.} is composed of audio and text from the LibriSpeech \cite{222panayotov2015librispeech} corpus. Librispeech, made up of audiobooks from the LibriVox project, was originally designed for ASR research and contains 982 hours of speech data from 2,484 speakers. The LibriTTS corpus inherits some of the properties of the LibriSpeech corpus, while addressing problems that make LibriSpeech less suitable for TTS tasks. For example, LibriTTS increases the sampling rate of audio files from 16 kHz to 24 kHz, splits speech at sentence breaks instead of at silences longer than 0.3 seconds, contains the original text and the standardized text, can extract contextual information (such as neighbouring sentences), and excludes utterances with significant background noise. The processed LibriTTS corpus consists of 585 hours of speech data at 24 kHz sampling rate from 2,456 speakers and its corresponding text transcripts.\par
There are other open source English corpora, such as the CMU ARCTIC corpus\footnote{The data is freely available for download from \url{http://www.festvox.org/cmu_arctic/}.} \cite{223kominek2003cmu} constructed by the Language Technologies Institute of Carnegie Mellon University for unit selection speech synthesis research. However, the amount of data in this corpus is too small to train the neural end-to-end TTS model well. Every year, The Blizzard Challenge, an international speech synthesis competition, provides participants with open source English speech data. For example, the corpus of The Blizzard Challenge 2011, 2013 and 2017\footnote{These data sets are freely available for download from\url{http://www.cstr.ed.ac.uk/projects/blizzard/} and can only be used for non-commercial use.} consists of tens of hours, hundreds of hours and 6 hours of audio and corresponding text transcripts of audiobooks read by a single speaker, with sampling frequencies of 16 kHz, 44.1 kHz and 44.1 kHz, respectively.
\subsection{Mandarin speech corpus}
Mandarin is the language with the largest number of speakers in the world, thus Mandarin TTS has also been widely researched and applied \cite{4gibiansky2017deep,6ping2017deep}. However, Mandarin has a complex tone and prosodic structure \cite{94minematsu2012improved}. Meanwhile, Chinese characters are ideograms, which are not directly related to pronunciation. It is necessary to convert the original Chinese text into phonemes or pinyin as audio transcription. Therefore, compared with English, the cost of recording and transcribing high-quality Mandarin corpus is higher, resulting in few open source high-quality Mandarin corpus. In order to facilitate researchers to conduct research on Mandarin TTS, several open source Mandarin corpora that can be used for TTS will be introduced.\par
CSMSC (Chinese Standard Mandarin Speech Copus)\footnote{The CSMSC corpus is available at \url{https://www.data-baker.com/open_source.html} for non-commercial use only.} \cite{224baker2017chinese} is a single-speaker Mandarin female voice corpus released by data-baker company. The corpus uses a professional recording studio and recording software for recording. The recording environment and equipment remain unchanged throughout the recording, and the signal-to-noise ratio (SNR) of the recording environment is not less than 35 dB. The audio format is a mono PCM WAV with a sampling frequency of 48 kHz 16 bit and an effective duration of approximately 12 hours. The recordings cover a variety of topics, such as news, fiction, technology, entertainment, dialogue, etc. The speech corpus is proofread, and rhythms and phoneme boundaries are manually edited.\par
AISHELL-3\footnote{The AISHELL-3 corpus is available at \url{http://www.aishelltech.com/aishell_3}, supporting academic research only and is prohibited from commercial use without permission.} \cite{217shi2020aishell} is a high-quality Mandarin corpus for multi-speaker TTS published by Shell Shell. It contains roughly 85 hours of emotion-neutral recordings spoken by 218 native Chinese mandarin speakers, as well as transcripts in Chinese character-level and pinyin-level. All utterances are recorded using a high-fidelity microphone (44.1 kHz, 16 bit) in a quiet indoor environment. The topics of the textual content spread a wide range of domains including smart home voice commands, news reports and geographic information.\par
DiDiSpeech\footnote{The DiDiSpeech data set is available for application on \url{https://outreach.didichuxing.com/research/opendata/}.} \cite{225guo2020didispeech} is a large open source Mandarin speech corpus released by DiDi Chuxing company. The corpus includes approximately 800 hours of speech data at a sampling rate of 48 kHz from 6,000 speakers and corresponding text transcripts. All speech data in the DiDiSpeech corpus are recorded in a quiet environment, and the audio with significant background noise is filtered. It is suitable for various speech processing tasks, such as voice conversion, multi-speaker TTS and ASR.
\subsection{Multilingual speech corpus}
There has been little research in the TTS field into languages other than English, partly because of the lack of available open source corpora. To enable TTS to be applied to more languages, some researchers have constructed speech corpora containing multiple languages, such as CSS10 \cite{226park2019css10} and Common Voice \cite{227livingstone2018ryerson}.\par
CSS10\footnote{The CSS10 corpus is available for free at \url{https://github.com/Kyubyong/CSS10}.} is a single-speaker corpus of ten languages, including Chinese, Dutch, French, Finnish, Japanese, Hungarian, Greek, German, Russian and Spanish. It is composed of short audio clips from LibriVox audiobooks and corresponding standardized transcriptions. All audio files are sampled at 22 kHz.\par
Common Voice\footnote{The Common Voice corpus is available for free at \url{https://commonvoice.mozilla.org/}.} is the largest public multilingual speech corpus, currently containing nearly 9,283 hours (7,335 hours verified) of speech data in 60 languages and fully open to the public. The project employs crowdsourcing for data collection and data validation. The audio clips are released as mono-channel, 16 bit MPEG-3 files with a 48 kHz sampling rate. This corpus is designed for ASR and rather noisy, thus denoising of the original audio data is required before it is used for the TTS task \cite{167nekvinda2020one}.
\subsection{Emotional speech corpus}
Emotional TTS has been extensively researched, but one of the problems currently in this field is the lack of publicly available emotional speech corpus and the difficulty of recording such data. None of the above-mentioned corpora contains explicit emotional information, and most of the existing emotional corpora cannot be effectively used to train the emotional TTS model based on deep learning, because these data sets contain a small number of sentences, such as RAVDESS \cite{227livingstone2018ryerson}, CREMA-D \cite{228cao2014crema}, GEEMP \cite{229banziger2012introducing}, EMO-DB \cite{230burkhardt2005database}, or contain noise, such as IMPROV \cite{231busso2016msp} and IEMOCAP \cite{232busso2008iemocap}.\par
To fill this gap, \citet{153tits2019emotional} released the Emov-DB corpus\footnote{The EmoV-DB database is available for free at \url{https://github.com/numediart/EmoV-DB}.}, which covers five emotions (amusement, anger, sleepiness, disgust, and neutral) and two languages (English and French). The English speech data is recorded by two male and two female speakers, and the French speech data is recorded by one male speaker. English sentences are taken from the CMU ARCTIC Corpus and French sentences from the SIWIS Corpus \cite{233honnet2017siwis}. Each audio file is recorded in 16 bits .wav format.
\section{Evaluation method}
\label{sec:6}
The speech quality is measured in three aspects: clarity, intelligibility and naturalness. However, at present, there is no uniform evaluation criterion for the quality of synthesized speech. In fact, different from the quantitative evaluation methods for tasks such as classification and prediction, since the final user is the audience, the level of generated speech quality often requires subjective qualitative evaluation. However, subjective evaluation is difficult to measure with strict standards, because there will be some deviations. In addition, some objective speech quality evaluation metrics also have reference value. Therefore, this section will summarize the evaluation methods of synthesized speech respectively from both subjective and objective aspects.
\subsection{Subjective evaluation method}
Subjective evaluation methods are usually more suitable for evaluating generative models, but they require significant resources and face challenges in the reliability, validity and reproducibility of results \cite{234ji2020comprehensive}. The most commonly used subjective evaluation method is the Mean Opinion Score (MOS), which measures naturalness by asking listeners to score the synthesized speech. MOS adopts a five-point scoring system, with higher scores indicating higher speech quality, which can be collected using the CrowdMOS toolkit \cite{235ribeiro2011crowdmos}. MUSHRA (Multiple Stimuli with Hidden Reference and Anchor) \cite{236recommendation20011534,237series2014method} is also a subjective listening test method. Specifically, the audio to be tested is mixed with natural speech as reference (upper limit) and total loss audio as anchor factor (lower limit). The listeners are asked to subjectively score the test audio, hidden reference audio and anchor factor through the double-blind listening test, with a score from 0 to 100. The 0-100 scale used by MUSHRA allows very small differences to be rated. The main advantage over the MOS methodology is that MUSHRA requires fewer participants to obtain statistically significant results.\par
All the above are absolute rating methods, and sometimes it is necessary to compare the speech quality generated by two models, which requires the use of relative rating methods, such as comparison mean opinion score (CMOS) and AB preference test. CMOS is used to compare the difference between the MOS value of the model under test and the baseline. AB preference test selects a better model or finds no significant difference between the two models by asking the listeners to compare the speech of the same sentence synthesized by the two models. The ABX preference test can be used when comparing multi-speaker TTS models or speech conversion models. Specifically, listeners are asked to listen to three speech fragments A, B and X respectively, where X represents the target speech, while A and B represent the speech generated by the two models respectively. The listeners are then asked to judge whether speech A or B is closer to X in terms of the personality characteristics of the speech, or can not give a clear judgment. Finally, the judgments of all listeners are counted to calculate the proportion of the speech synthesized by each model that sounded more like the target speech.
\subsection{Objective evaluation method}
The objective evaluation method is mainly the quantitative evaluation of the TTS model and the generated speech. The differences between the generated samples and the real samples are usually used to evaluate the model. However, these evaluation metrics can only reflect the data processing ability of the model to a certain extent, and cannot truly reflect the quality of the generated speech.\par
The most intuitive way to objectively evaluate the prosody and accuracy of synthesized speech is to directly calculate the root mean square error (RMSE), absolute error and negative log likelihood (NLL) of $f_0$, pitch, $c_0$ (the 0-th cepstrum coefficient) and duration of reference audio and predicted audio, as well as the character error rate (CER), word error rate (WER) and utterance error rate (UER) of synthesized speech.
\par
Another commonly used objective evaluation metric for judging the difference between the generated samples and the real samples is Mel-Cepstral Distortion (MCD) \cite{238kubichek1993mel}. MCD quantifies the reconstruction performance of Mel-Frequency Cepstrum Coefficients (MFCC) by calculating the spectral distance between synthesized and reference mel-spectral features. Its calculation formula is:
\begin{equation}
	MCD_K=\frac{1}{T}\sum_{t=0}^{T-1}\sqrt{\sum_{k=1}^{K}(c_{t,k}-c_{t,k}^{\prime})^2}
\end{equation}
where $c_{t,k}$ and $c_{t,k}^{\prime}$ are the k-th MFCC from the t-th frame of the reference and predicted audio, respectively. The MCD is usually calculated using the mean square error (MSE) calculated by the MFCC features of $K=13$ dimensions. The lower the value of MCD, the higher the quality of synthesized speech. It can be used to evaluate timbral distortion, and its unit is db. A similar evaluation metric is mel-spectral distortion (MSD). MSD is calculated in the same way as MCD, but it is calculated with the logarithmic mel-spectral amplitude rather than cepstrum coefficient, which captures the harmonic content not found in MCD.\par
Gross Pitch Error (GPE) and Voicing Decision Error (VDE) are two commonly used metrics to measure the error rate of synthesized speech \cite{240nakatani2008method}. GPE is the estimation error of the audio $f_0$ value, defined as \cite{33skerry2018towards}:
\begin{equation}
	GPE=\frac{\sum_t1[\mid p_t-p_t^{\prime}\mid>0.2p_t]1[v_t]1[v_t^{\prime}]}{\sum_t1[v_t]1[v_t^{\prime}]}
\end{equation}
where $p_t$, $p_t^{\prime}$ are the pitch signals from the reference and predicted audio, $v_t$, $v_t^{\prime}$ are the voicing decisions from the reference audio and predicted audio, and $1$ is the indicator function. The GPE measures the percentage of voiced frames in the predicted audio that deviate in pitch by more than $20\%$ compared to the reference. VDE is defined as \cite{33skerry2018towards}:
\begin{equation}
	VDE=\frac{\sum_{t=0}^{T-1}1[v_t\not=v_t^{\prime}]}{T}
\end{equation}
where $v_t$, $v_t^{\prime}$ are the voicing decisions of the reference and predicted audio, $T$ is the total number of frames, and $1$ is the indicator function. VDE is used to measure the frame-level voicing decision error rate of the predicted audio. The lower these two metrics, the better. However, some algorithms have low GPE but high VDE. In order to reduce the values of VDE and GPE at the same time, \citet{239chu2009reducing} combined GPE and VDE and proposed $f_0$ Frame Error (FFE) metric. FFE is used to measure the percentage of frames that either contain a $20\%$ pitch error (according to GPE) or a voicing decision error (according to VDE), defined as \cite{33skerry2018towards}:
\begin{equation}
\begin{split}
	& FFE= \\
	& \frac{\sum_{t=0}^{T-1}(1[\mid p_t-p_t^{\prime}\mid>0.2p_t]1[v_t]1[v_t^{\prime}]+1[v_t\not=v_t^{\prime}])}{T} 
\end{split}
\end{equation}
FFE is used to calculate the ratio of the difference between the predicted pitch and the true pitch, which can quantify the reconstruction error of the $f_0$ trajectory. The lower the value, the better.\par
\citet{113binkowski2019high} also proposed four metrics for evaluating TTS models: unconditional and conditional Fréchet DeepSpeech Distance (FDSD, cFDSD) and Kernel DeepSpeech Distance (KDSD, cKDSD). These metrics are inspired by the commonly used metrics for evaluating GAN-based image generation models \cite{242heusel2017gans,243binkowski2018demystifying}, which judge the quality of the synthesized speech by calculating the distance between the synthesized audio and the reference audio. Moreover, the quality of synthesized speech waveform can also be evaluated by calculating Perceptual evaluation of speech quality (PESQ) \cite{241rix2001perceptual} of reference speech and synthesized speech, with the higher the value, the better.
\section{Future development direction}
\label{sec:7}
With the development of deep learning fields such as NMT, ASR, image generation and music generation, although existing TTS methods can synthesize high-fidelity speech by drawing on various Seq2Seq models and generation models, they still have many shortcomings. For example, the existing TTS technology based on deep learning is still unable to stably synthesize speech in real time, and the quality of the generated speech is difficult to be guaranteed. For example, the end-to-end TTS technology based on deep learning has not been able to synthesize speech stably in real time, and the quality of the generated speech cannot be guaranteed. Therefore, a large part of TTS models currently used in the industry are based on waveform cascade technology \cite{1capes2017siri}. Moreover, the state-of-the-art TTS technology is limited to a few common languages such as English and Mandarin. Since it is difficult to obtain the data pairs of $\langle$text, speech$\rangle$, there has been little research on minor languages and dialects.\par
Based on the above introduction and summary of TTS method, it can be concluded that there will be at least the following development directions in the field of TTS in the future:
\begin{itemize} 
	\item[$\bullet$]\emph{Control the style of speech in a precise and fine-grained manner} Speaking styles such as emotion, intonation and rhythm often change during conversation. However, current neural TTS systems cannot precisely control these style features of speech individually. How to achieve fine-grained style control of speech at word level and phrase level will also be the focus of TTS research in the future. In addition, due to the difficulty in recording and labeling emotional speech data, how to effectively use emotional speech data limited in quantity and quality to train the TTS model and enable it to learn the representation methods of various style features in speech is also an urgent problem in the field of TTS.
	\item[$\bullet$]\emph{In-depth research on the representation method of speech signal in deep neural network} Children learn to speak long before they learn to read and write. They can conduct a dialogue and produce novel sentences, without being trained on an annotated corpus of speech and text or aligned phonetic symbols. Presumably, they achieve this by recoding the input speech in their own internal phonetic representations (proto-phonemes or proto-text) \cite{270dunbar2019zero}. This idea can also be applied to TTS systems, as stated in the goal of the ZeroSpeech Challenge: extract acoustic units from speech signals by unsupervised learning and create good data representation. Therefore, representation learning and meta-learning can be used to improve the modeling ability and learning efficiency of TTS model for speech data, thus greatly reducing the labeled speech data required for training.
	\item[$\bullet$]\emph{Build a fully end-to-end TTS pipeline} Although the existing TTS models are all called end-to-end, most of them are divided into three parts: text front-end, acoustic model and vocoder. These three modules need to be trained separately, and the errors generated by each module will gradually accumulate. The latest TTS frameworks such as ClariNet \cite{25ping2018clarinet}, FastSpeech 2s \cite{15ren2020fastspeech}, EATS \cite{114donahue2020end} and Wave-Tacotron combine these modules and claim to be fully end-to-end for training and inference. However, they still generate intermediate acoustic features as the condition of the audio generation module, essentially similar to other methods. A fully end-to-end model that maps original text or phonemes directly to speech waveforms would greatly simplify the TTS pipeline.
	\item[$\bullet$]\emph{Apply the deep learning methods used in other tasks to TTS} First, as a generation task, speech synthesis and image generation have great similaritie. Many methods used in TTS are inspired by image generation methods. For example, MelNet \cite{252vasquez2019melnet} regards the speech spectrogram as an image, and synthesizes the mel-spectrogram using a 2-D multi-scale autoregressive generation method. The methods of generating images and speech with specific styles are also very similar. Second, the alignment method in the acoustic model can learn from the methods in NMT and ASR, which are also Seq2Seq models. Third, as recognition and generation are dual tasks, multi-task learning can be adopted to combine recognition and generation models to improve each other and reduce the demand for labeled data during training. In addition to combining TTS and ASR \cite{57tjandra2017listening,58liu2018improving,68ren2019almost,245tjandra2018machine,264xu2020lrspeech}, it is also possible to combine speaker recognition with multi-speaker TTS \cite{131chen2018sample,245tjandra2018machine}, and combine speech emotion recognition with emotional speech synthesis \cite{184li2021controllable} for dual training.
\end{itemize}
\section{Conclusion}
\label{sec:8}
The research of end-to-end TTS technology based on deep learning has become a hot topic in the field of artificial intelligence. In order to make researchers to have a clear understanding of the latest TTS paradigm, this paper summarizes the latest technologies used in each module of the TTS system in detail, and classifies the methods according to their characteristics and compares their advantages and disadvantages. Furthermore, the public speech corpus for various TTS tasks and the commonly used subjective and objective speech quality evaluation methods are also summarized. Finally, some suggestions for the future development direction of TTS are put forward.

%\begin{acknowledgements}
%\end{acknowledgements}


\bibliographystyle{spbasic}      % basic style, author-year citations
%\bibliographystyle{plain}
\bibliography{refer}
%\begin{thebibliography}{}
% etc
%\end{thebibliography}

\end{document}